\documentclass[a4paper,11pt]{article}
\usepackage{jheppub} 
\usepackage{lineno}

\arxivnumber{2605.28752} 

\title{Inflation with vector fields revisited: non-Gaussianities}

\author{Chong-Bin Chen}
\affiliation{Department of Physics, Nanchang University, Nanchang, 330031, China}
\affiliation{Center for Relativistic Astrophysics and High Energy Physics,
Nanchang University, Nanchang 330031, China}

\emailAdd{chongbin@ncu.edu.cn}

\abstract{We revisit the resulting bispectrum of inflation with kinetic-coupled vector fields by organizing the dynamics in terms of $h$, which measures the vector kinetic contribution relative to that of the scalar field. We derive a low-energy effective field theory (EFT) and evaluate the bispectrum of the curvature perturbation for the large-$h$ regime. For $h\gg1$, the entropic perturbation becomes heavy and can be integrated out; the remaining curvature mode has an imaginary sound speed and undergoes transient growth before horizon crossing. In contrast to the $h\ll1$ regime, where transfer from the vector sector persists outside the horizon and produces a local-type contribution enhanced as $h^2N_K^3$, we find that the bispectrum signals are controlled by the matching parameter $x\simeq1.63Ah$, where $A<1$ is a numerical factor remaining dependent on the UV-to-EFT matching prescription. As a result, in addition to the known flattened-enhanced signals scaling as $x^3$, a flattened-enhanced signal scaling as $x^2$ and a pronounced local projection scaling as $x$ are present. The bispectrum therefore distinguishes vector-supported inflationary dynamics even for an exactly isotropic background.}

\begin{document}
\maketitle
\flushbottom

\section{Introduction}
Inflationary models with vector fields provide a controlled setting in which additional degrees of freedom can modify primordial perturbations \cite{Ratra:1991bn,Turner:1987bw,Dvali:2007ks,Martin:2007ue,Demozzi:2009fu,Durrer:2024ibi,Valenzuela-Toledo:2009yve,Valenzuela-Toledo:2009bzd,BeltranAlmeida:2011db,Rodriguez:2013cj,Choi:2015wva}. When the kinetic function of the vector fields depends on the inflaton (i.e. $f_{ab}(\phi)F^{a}_{\mu\nu}F^{b\mu\nu}$), the vector energy density can remain relevant during slow roll and feed back into both the background motion and the fluctuation dynamics \cite{Watanabe:2009ct, Watanabe:2010fh, Kanno:2010nr, Yamamoto:2012tq, Maleknejad:2012fw, Bartolo:2012sd,Himmetoglu:2009mk,Gumrukcuoglu:2010yc,Gumrukcuoglu:2007bx,Emami:2013bk,Bamba:2008ja}. The idea of introducing additional fields is not only aimed at solving theoretical problems; studying the non-Gaussianities they generate provides an important window into microscopic physics beyond the Hubble scale \cite{Wands:2010af, Chen:2010xka, Wang:2013eqj, Renaux-Petel:2015mcb}.

The phenomenology is governed by the relative importance of the vector and scalar kinetic sectors. In this paper this ratio is parametrized by $h$, defined below in terms of $\sqrt{\epsilon_A/2\epsilon_\phi}$. For $h\ll1$, vector perturbations act as weakly coupled additional modes. Their transfer to the curvature perturbation continues on superhorizon scales and generates a local-type bispectrum with an amplitude enhanced by the duration of inflation after horizon exit \cite{Watanabe:2010fh,Bartolo:2012sd,Gorji:2020vnh,Yamamoto:2012sq,Funakoshi:2012ym}. This cumulative behavior places a stringent constraint on the weak-vector regime. The constraint implies the presence of a tiny but non-zero parameter, which may give rise to fine-tuning challenge \cite{Naruko:2014bxa,Fujita:2017lfu}.

The situation changes qualitatively when $h\gtrsim1$. However, in this parameter regime the vector fields cannot be treated as weakly coupled modes, so that the perturbative method  fails \cite{Funakoshi:2012ym}. Fortunately, \cite{Chen:2022ccf,Chen:2023bcz,Chen:2025qyv} found that in the $h\gg1$ regime, the entropic fluctuation becomes heavy and can be integrated out at low energies, leaving an effective single-mode description for the curvature perturbation \cite{Cheung:2007st, Achucarro:2010da, Achucarro:2012yr,Baumann:2011su}. The modes contributing to the non-Gaussianities are quite different from those in the $h\ll1$ case, and hence we need to revisit it. The resulting effective theory has an imaginary sound speed. An effective field theory with imaginary sound speed was studied in multi-field inflation with hyperbolic field space \cite{Brown:2017ena, Mizuno:2017idt, Bjorkmo:2019fia,Renaux-Petel:2015mca, Garcia-Saenz:2018qfq, Renaux-Petel:2017cet,Garcia-Saenz:2018ifx, Fumagalli:2019noh,Garcia-Saenz:2019njm}. Such a sound speed signals transient exponential evolution before horizon crossing, and reorganizes the bispectrum into interactions whose momentum dependence can be calculated explicitly. In particular, we find that flattened configurations are amplified by the interference between growing and decaying modes, while interactions specific to the vector-field realization also generate a sizable local component.

We choose an isotropic configuration of vector fields as a toy model to preserve isotropy and computational simplicity. An isotropic configuration can be realized by a triad of vector fields, or effectively by a suitable ensemble \cite{Bento:1992wy,Golovnev:2008cf}, so that the vector sector carries a non-vanishing background energy density without selecting a preferred spatial direction \cite{Yamamoto:2012sq,Funakoshi:2012ym,Firouzjahi:2018wlp,Gorji:2020vnh,Chen:2022ccf,Chen:2023bcz}. We should emphasize that our discussion can be expanded to one vector-field theory \cite{Chen:2025qyv}, where a large anisotropy is inevitable. We compute the corresponding bispectrum in both limits of $h$. Our emphasis is on the large-$h$ regime. We determine their correlations with standard non-Gaussianities templates and show how their competition changes the dominant projected amplitude as the EFT matching scale, which is proportion to $h$, is varied.

The paper is organized as follows. We first present the isotropic background solution and define the parameter $h$. We then derive the quadratic and cubic actions for scalar perturbations and obtain the low-energy effective action in the large-$h$ regime. The bispectrum is computed next for small and large $h$, followed by a template analysis of the large-$h$ non-Gaussianities. Technical details of the perturbative action and Hamiltonian are collected in the appendix.

\section{Background}
We choose the isotropic and homogeneous configuration of the vector fields as
\begin{align}
    A^a_{\ 0}=0,\ \ \ \ \ A^a_{\ i}=\mathbb{A}\delta_{ai}, \ \ \ \ \ f_{ab}=f^2\delta_{ab}
\end{align}
so that the Universe remains isotropic and homogeneous at the background level,
\begin{align}
    ds^2=-dt^2+a(t)^2\delta_{ij}dx^idx^j,
\end{align}
where $t$ is cosmic time and $a(t)$ is the scale factor. Such a vector configuration can arise from the leading contribution of many randomly oriented vector fields \cite{Bento:1992wy} or from a non-Abelian gauge field when the gauge coupling can be neglected \cite{Murata:2011wv}. In the following we keep the discussion general rather than committing to either realization. With this configuration, the background equations of motion are
\begin{align}
     &M_{\text{pl}}^2H^2\left(3-\epsilon_{\phi}-\frac{3}{2}\epsilon_A\right)=V(\phi),\ \ \ \ \ \ \ \epsilon=\epsilon_{\phi}+\epsilon_A,
\end{align}
where the slow-roll parameters are defined by
\begin{align}
    \epsilon\equiv-\frac{\dot{H}}{H^2},\ \ \ \ \ \ \ \epsilon_{\phi}\equiv\frac{\dot{\phi}^2}{2M_{\text{pl}}^2H^2},\ \ \ \ \ \ \ \epsilon_A\equiv \frac{f^2\dot{\mathbb{A}}^2}{a^2M_{\text{pl}}^2H^2}.
\end{align}
The equations of motion of the scalar field and the vector fields are given by
\begin{align}
    &\ddot{\phi}+3H\dot{\phi}+V_{\phi}-3\frac{f_{\phi}}{f}\epsilon_AM_{\text{pl}}^2H^2=0,\ \ \ \ \ \ \ \frac{d}{dt}(af\dot{\mathbb{A}})=0, \label{eq}
\end{align}
where $V_{\phi}\equiv\partial V/\partial \phi$ and $f_{\phi}\equiv\partial f/\partial \phi$. Without loss of generality, we assume $\dot{\phi}<0$ and $\dot{\mathbb{A}}>0$. The last term in the equation for $\phi$ arises from its coupling to the vector fields. A suitable choice of $f$ sustains this contribution and consequently modifies the inflaton dynamics.

We define a parameter
\begin{align}
     h\equiv \sqrt{\frac{\epsilon_A}{2\epsilon_{\phi}}},
\end{align}
which measures the vector-field energy density relative to the scalar kinetic energy and controls the perturbation theory. Although $\epsilon_A,\epsilon_{\phi}\ll1$ during slow-roll inflation, $h$ need not be small when $\epsilon_A\gtrsim\epsilon_{\phi}$. An example is the anisotropic inflationary attractor \cite{Watanabe:2009ct, Watanabe:2010fh, Kanno:2010nr, Yamamoto:2012tq} (which also has an isotropic counterpart \cite{Yamamoto:2012sq,Funakoshi:2012ym,Firouzjahi:2018wlp,Gorji:2020vnh,Chen:2022ccf,Chen:2023bcz}), for which $\epsilon_A\simeq\text{constant}$ at leading order in slow roll. A broad region of parameter space yields attractors with $h\gtrsim1$ \cite{Chen:2021nkf,Chen:2022ccf}. Throughout this paper we take $h$ to be constant.

A relation useful for the perturbation analysis follows by taking the time derivative of the Friedmann equation
and combining it with the equation of motion for $\phi$:
\begin{equation}\label{eq2}
    \frac{f_{\phi}}{f}M_{\text{pl}}\sqrt{2\epsilon_{\phi}}=2\left(1+\frac{1}{4}\eta_{A}-\frac{1}{2}\epsilon\right).
\end{equation}
where $\eta_{A}\equiv\dot{\epsilon}_A/(H\epsilon_A)$. Substitution into the equation of motion for $\phi$ gives $3H\dot{\phi}\simeq-V_{\phi}/(1+2h^2)$ in the slow-roll approximation. Thus part of the potential energy sources the vector-field kinetic energy and alters the background inflaton motion.

\section{Perturbation}
\subsection{Quadratic and cubic action}
We derive the perturbative action up to third order. Because the background is isotropic, perturbations can be separated into scalar, vector, and tensor sectors. The perturbed matter fields are written as
\begin{align}
    \phi=\phi(t)+\delta\phi,\ \ \ \ \ \ A^a_{\ 0}=\partial_a\mathbb{Y},\ \ \ \ \ \ 
A^a_{\ i}=\left(\mathbb{A}+\delta\mathbb{A}\right)\delta_{ai}+\epsilon_{iab}\partial_b\mathbb{U}+\partial_i\partial_a\mathbb{M}.
\end{align}
We consider only the scalar sector. The vector fields possess the gauge freedom $A^a_{\ \mu}\rightarrow A^a_{\ \mu}+\partial_{\mu}\rho_a$, where $\rho^a$ is arbitrary. Decomposing it as $\rho_a=\partial_a\rho+\hat{\rho}_a$, the scalar perturbations transform as
\begin{align}
    \mathbb{Y}\to\mathbb{Y}+\dot{\rho},\ \ \ \ \ \ \ \mathbb{M}\to\mathbb{M}+\rho
\end{align}
under a gauge transformation. We fix this redundancy by setting $\mathbb{M}=0$, which we refer to as the spatially-flat gauge for the vector perturbations. The perturbed field strengths are then
\begin{align}
    &F^{(a)}_{0i}=\dot{\mathbb{A}}\delta_{ai},\ \ \ \ F^{(a)}_{ij}=0,\ \ \ \ \delta F^{(a)}_{0i}=\delta\dot{\mathbb{A}}\delta_{ai}+\epsilon_{iab}\partial_b\dot{\mathbb{U}}-\partial_i\partial_a\mathbb{Y},\nonumber\\
    &\delta F^{(a)}_{ij}=\partial_i\delta\mathbb{A}\delta_{aj}-\partial_j\delta\mathbb{A}\delta_{ai}+\epsilon_{jab}\partial_b\partial_i\mathbb{U}-\epsilon_{iac}\partial_c\partial_j\mathbb{U}.
\end{align}

The scalar perturbations of gravity can also be included. In the spatially-flat gauge, however, the dominant contributions to the observable quantities arise from the matter perturbations $f^2F^2$ in the slow-roll approximation \cite{Watanabe:2010fh,Bartolo:2012sd,Emami:2013bk}. We therefore neglect gravitational backreaction in the following analysis.

We define a variable $\delta Q$ as
\begin{align}\label{eq4}
    \delta Q\equiv\frac{\sqrt{2}f}{a}\delta\mathbb{A}.
\end{align}
We focus on the case $\epsilon_A\simeq\text{const.}$, for which $f\propto a^{-2}$. Consequently, $\dot{f}/f\simeq-2H$ and $\delta\dot{\mathbb{A}}=(a/f)(\delta\dot{Q}+3H\delta Q)$. The perturbation $\mathbb{Y}$ is non-dynamical and can be eliminated through its constraint equation in the quadratic action (see appendix \ref{appendix}). Combining these relations with (\ref{eq2}), we obtain the quadratic and cubic Lagrangians
\begin{align}
    \mathcal{L}^{(2)}\simeq{}&\frac{a^3}{2}\bigg[\delta\dot{\phi}^2-\frac{1}{a^2}(\partial_i\delta\phi)^2+8h^2H^2\delta\phi^2
+\delta \dot{Q}^2-\frac{1}{a^2}(\partial_i\delta Q)^2+8\sqrt{2}hH\delta\phi\delta \dot{Q}\nonumber\\
    &+24\sqrt{2}hH^2\delta\phi\delta Q+\frac{2f^2}{a^2}(\partial_i\dot{\mathbb{U}})^2-\frac{2f^2}{a^4}(\partial_i\partial_j\mathbb{U})^2\bigg],\\
    \mathcal{L}^{(3)}\simeq{}&\frac{a^3}{M_{\text{pl}}\sqrt{2\epsilon_\phi}}\bigg[-\frac{16h^2}{3}H^2\delta\phi^3+\frac{8\sqrt{2}h}{3}H\left(\delta\dot{Q}+3H\delta Q\right)\delta\phi^2+\frac{4}{3}\left(\delta\dot{Q}+3H\delta Q\right)^2\delta\phi\nonumber\\
    &-\frac{2}{a^2}(\partial_i\delta Q)^2\delta\phi+\frac{4f^2}{a^2}\left((\partial_i \dot{\mathbb{U}})^2-\frac{1}{2a^2}(\partial_i\partial_j\mathbb{U})^2-\frac{1}{2a^2}(\partial^2\mathbb{U})^2\right)\delta\phi\bigg].\label{L3phiQ}
\end{align}
At quadratic order, $\mathbb{U}$ decouples from the other perturbations because it represents the magnetic component of the vector fields, while the isotropic background contains no magnetic field. Although $\mathbb{U}$ couples to $\delta\phi$ nonlinearly, these interactions do not give an important contribution to the curvature bispectrum.

\subsection{Theory of the curvature perturbation}
The variable most directly related to observables is the curvature perturbation. Since the vector-field energy density contributes to the inflationary background, it also contributes to the curvature fluctuation. The comoving curvature perturbation is defined by $\mathcal{R}=\psi-H\delta u$, where the velocity perturbation obeys $\delta T^0_{\ i}=(\rho+p)\partial_i\delta u$. For non-vanishing background vector fields, this velocity receives contributions from both the scalar field and the vector fields. In the spatially-flat gauge, one obtains
\cite{Gorji:2020vnh,Chen:2022ccf}
\begin{align}\label{R}
    \mathcal{R}=H\frac{a^2\dot{\phi}\delta\phi+2f^2\dot{\mathbb{A}}\delta\mathbb{A}}{a^2\dot{\phi}^2+2f^2\dot{\mathbb{A}}^2}.
\end{align}
The magnetic perturbation $\mathbb{U}$ does not enter the curvature perturbation because the magnetic background vanishes. We define the reduced curvature perturbation by $\mathcal{R}_c\equiv\sqrt{2\epsilon}M_{\text{pl}}\mathcal{R}$.

The relative weights of the vector-field and inflaton perturbations become transparent after rewriting the curvature perturbation in terms of $h$:
\begin{align}
    \mathcal{R}_c\equiv\frac{-\delta\phi+\sqrt{2}h\delta Q}{\sqrt{1+2h^2}}.
\end{align}
Their weight ratio is therefore $\sqrt{2}h$. In analogy with multi-field inflation, we introduce an angle $\mathbb{\vartheta}$ such that $\mathcal{R}_c=\cos\mathbb{\vartheta}\ \delta\phi+\sin\mathbb{\vartheta}\ \delta Q$, and define an ``entropic'' perturbation by
\begin{align}
    \mathcal{F}\equiv -\frac{\sqrt{2}h\delta\phi+\delta Q}{\sqrt{1+2h^2}}.
\end{align}
Using $\epsilon=(1+2h^2)\epsilon_\phi$, the quadratic and cubic Lagrangians can then be expressed in terms of the curvature and entropic perturbations as
\begin{align}\label{L2}
    \mathcal{L}^{(2)}={}&\frac{a^3}{2}\bigg[\dot{\mathcal{R}}_c^2-\frac{1}{a^2}(\partial_i \mathcal{R}_c)^2 -m_{\sigma}^2\mathcal{R}_c^2+\dot{\mathcal{F}}^2-\frac{1}{a^2}(\partial_i\mathcal{F})^2-m_s^2\mathcal{F}^2\nonumber\\
    &-2\sqrt{2}hH\left(4\dot{\mathcal{R}}_c
    +\frac{16h^2}{1+2h^2}H\mathcal{R}_c\right)\mathcal{F}+\frac{2f^2}{a^2}(\partial_i\dot{\mathbb{U}})^2-\frac{2f^2}{a^4}(\partial_i\partial_j\mathbb{U})^2\bigg]
\end{align}
and
\begin{align}\label{L3noApp}
    \mathcal{L}^{(3)}={}&\frac{a^3}{M_{\text{pl}}\sqrt{\epsilon}}\bigg[\frac{4h(8h^4-12h^2-9)}{3(1+2h^2)}H^2\mathcal{F}^3+\frac{2\sqrt{2}(16h^4+4h^2-3)}{1+2h^2}H^2\mathcal{F}^2\mathcal{R}_c\nonumber\\
    &+\frac{8h(2+3h^2)}{1+2h^2}H^2\mathcal{F}\mathcal{R}_c^2-\frac{4\sqrt{2}h^2}{3(1+2h^2)}H^2\mathcal{R}_c^3+\frac{8\sqrt{2}(2h^2+3)h^2}{3(1+2h^2)}H\mathcal{F}^2\dot{\mathcal{R}}_c\nonumber\\
    &+\frac{8h(3-2h^2)}{3(1+2h^2)}H\mathcal{F}\mathcal{R}_c\dot{\mathcal{R}}_c-\frac{16\sqrt{2}h^2}{3(1+2h^2)}H\mathcal{R}_c^2\dot{\mathcal{R}}_c-\frac{8h^3}{3(1+2h^2)}\mathcal{F}\dot{\mathcal{R}}_c^2\nonumber\\
    &-\frac{4\sqrt{2}h^2}{3(1+2h^2)}\mathcal{R}_c\dot{\mathcal{R}}_c^2+\frac{2h}{a^2(1+2h^2)}\mathcal{F}(\partial_i\mathcal{F})^2-\frac{4\sqrt{2}h^2}{a^2(1+2h^2)}\mathcal{F}\partial_i\mathcal{F}\partial_i\mathcal{R}_c\nonumber\\
    &+\frac{4h^3}{a^2(1+2h^2)}\mathcal{F}(\partial_i\mathcal{R}_c)^2+\frac{\sqrt{2}}{a^2(1+2h^2)}\mathcal{R}_c(\partial_i\mathcal{F})^2-\frac{4h}{a^2(1+2h^2)}\mathcal{R}_c\partial_i\mathcal{R}_c\partial_i\mathcal{F}\nonumber\\
    &+\frac{2\sqrt{2}h^2}{a^2(1+2h^2)}\mathcal{R}_c(\partial_i\mathcal{R}_c)^2-2\sqrt{2}{}\Xi_{\mathbb{U}}\left(\mathcal{R}_c+\sqrt{2}h\mathcal{F}\right)-  \frac{4h}{3(1+2h^2)}\mathcal{F} \dot{\mathcal{F}}^2\nonumber\\
    &- \frac{2\sqrt{2}}{3(1+2h^2)}\mathcal{R}_c \dot{\mathcal{F}}^2-\frac{8h(3+2h^2)}{3(1+2h^2)}H\mathcal{F}^2\dot{\mathcal{F}} + \frac{4\sqrt{2}(2h^2-3)}{3(1+2h^2)}H\mathcal{F}\mathcal{R}_c\dot{\mathcal{F}}\nonumber\\
    &+ \frac{16h}{3(1+2h^2)}H\mathcal{R}_c^2\dot{\mathcal{F}} + \frac{8\sqrt{2}h^2}{3(1+2h^2)}\mathcal{F}\dot{\mathcal{R}}_c\dot{\mathcal{F}}+ \frac{8h}{3(1+2h^2)}\mathcal{R}_c^2\dot{\mathcal{F}}\bigg]
\end{align}
where the masses of two perturbations and $\Xi_U$ are 
\begin{align}
    &m_{\sigma}^2\equiv\frac{16h^2}{1+2h^2}H^2,\ \ \ \ \ \ \ \ m_s^2= -\frac{8h^2\left(3+2h^2\right)}{1+2h^2}H^2,\\
    &\Xi_{\mathbb{U}}\equiv \frac{f^2}{a^2}\left((\partial_i \dot{\mathbb{U}})^2-\frac{1}{2a^2}(\partial_i\partial_j\mathbb{U})^2-\frac{1}{2a^2}(\partial^2\mathbb{U})^2\right).
\end{align}
Both masses depend on $h$. For $h\ll1$, the two perturbations are nearly massless. For $h\gg1$, instead, $m_{\sigma}^2\sim H^2$ and $|m_{s}^2|\sim h^2H^2\gg H^2$, so that the entropic perturbation is heavy. Such a high-frequency degree of freedom can be integrated out at low energies \cite{Achucarro:2010da}. The low-energy EFT for $h\gg1$ in this model was discussed in \cite{Chen:2023bcz}; below we study its non-Gaussianities in Sec.~\ref{largeh}.

\subsubsection{Small $h$}
We can expand the cubic Lagrangian near $h=0$ up to $h^2$ as
\begin{align}\label{L_smallh}
    \mathcal{L}^{(3)}_{h\ll1}\simeq{}\frac{a^3}{M_{\text{pl}}\sqrt{2\epsilon}}\bigg[\mathcal{L}^{(3)}_0+h\mathcal{L}^{(3)}_1+h^2\mathcal{L}^{(3)}_2+\mathcal{O}(h^3)\bigg],
\end{align}
where $\mathcal{L}^{(3)}_i$ are given by (\ref{L3_0})--(\ref{L3_2}). The expansion contains many interactions, but only a subset contributes at leading order to the total non-Gaussianities. As studied in \cite{Gorji:2020vnh,Yamamoto:2012sq}, observational constraints in this regime force $h$ to be very small; terms at higher order in $h$ can therefore be neglected.

\subsubsection{Large $h$}\label{largeh}
\begin{figure}[t]
\centering
\includegraphics[scale=0.6]{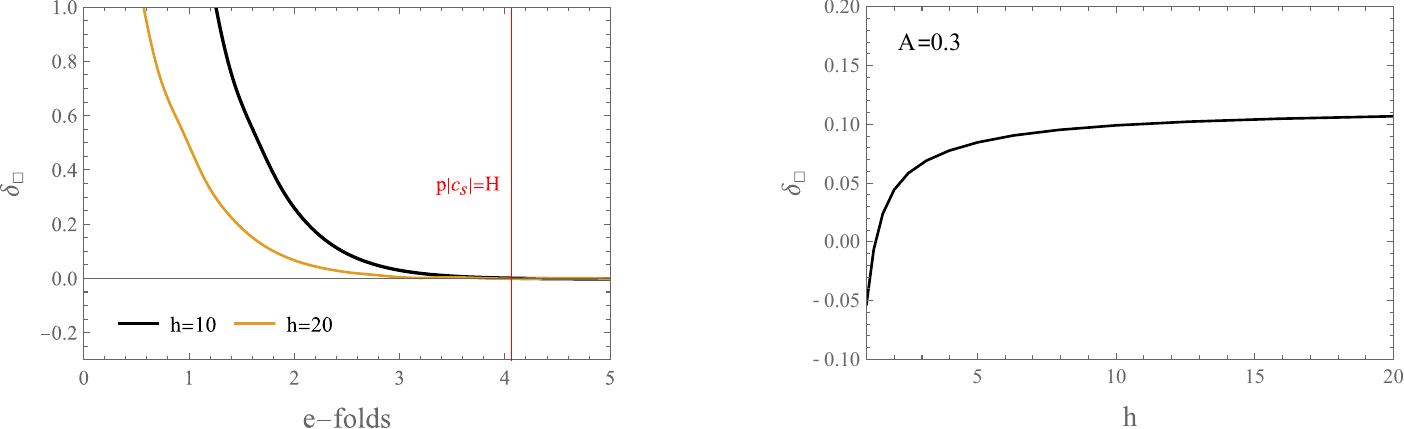}
\caption{\label{fig:delta}Validity of the low-energy EFT in the large-$h$ regime.
\textit{Left:} Evolution of the higher-derivative correction 
$\delta_{\Box}$ as a function of the number of $e$-folds for 
$h=10$ and $h=20$. \textit{Right:} $\delta_{\Box}$ evaluated at the EFT matching scale 
$p_i=A|m_s|$ as a function of $h$, with $A=0.3$.}
\end{figure}

Our main interest is the regime $h\gg1$, where $|m_s|\gg H$. Defining $\square\equiv-\partial^2/\partial t^2-3H\partial/\partial t-k^2/a^2$, we expand the propagator as $1/(m^2_s-\square)=1/m^2_s+\square/m^4_s+\cdots$ in the equation of motion. The leading-order solution for $\mathcal{F}$ is \cite{Chen:2023bcz}
\begin{align}\label{FLO}
    \mathcal{F}_{\text{LO}}=-\frac{\sqrt{2}hH}{m_s^2}\left(4\dot{\mathcal{R}}_c
    +\frac{16h^2}{1+2h^2}H\mathcal{R}_c\right).
\end{align}
We shall briefly examine the regime of validity of our approximations, namely, for what range of $h$ and down to what redshifted physical momentum 
$p$ a low-energy effective theory EFT can be derived. The next-to-leading order correction of $\mathcal{F}$ is
\begin{align}\label{F_corr}
    \mathcal{F}=\mathcal{F}_{\rm LO}\left[1+\delta_{\square}+\mathcal{O}\left(\frac{\square^2}{m_s^4}\right)\right],\ \ \ \ \ \delta_{\square}\equiv\frac{1}{m_s^2}\frac{\square\mathcal{F}_{\rm LO}}{\mathcal{F}_{\rm LO}}.
\end{align}
In FIG. \ref{fig:delta}(left), we show the evolution of $\delta_{\square}$ as a function of time near horizon crossing for $h=10$ and $h=20$. The larger $h$ is, the earlier the theory enters the EFT-valid regime (i.e., $\delta_{\square}\lesssim1$), and during this period $\delta_{\square}$ decays rapidly as the momentum redshifts.

We define the physical momentum at which the EFT becomes effective as $p_i$. The validity of EFT is equivalent to requiring $p_i^2/|m_s|^2\ll 1$. We can use $p_i=A|m_s|$ to denote this cutoff, where $A$ is a numerical factor less than unity. In FIG. \ref{fig:delta}(right), we show $\delta_{\square}$ for different values of $h$ with $A=0.3$. It can be seen that the correction stabilizes for $h\gtrsim5$, meaning that the EFT description is sufficiently valid in this regime. For $A=0.3$, the correction is approximately stable at about $10\%$.

Substituting the leading order solution of $\mathcal{F}$ into the quadratic Lagrangian, we obtain the effective Lagrangian of $\mathcal{R}$:
\begin{align}
    \mathcal{L}^{(2)}_{\text{eff}}=\frac{a^3}{2c_s^2}\bigg[\dot{\mathcal{R}}_c^2-\frac{c_s^2}{a^2}(\partial_i \mathcal{R}_c)^2\bigg],
\end{align}
where the speed of sound is 
\begin{align}\label{cs}
    \frac{1}{c_s^2}= 1-\frac{4\left(1+2h^2\right)}{3+2h^2}.
\end{align}
The equation of motion of $\mathcal{R}_c$ is performed as $\ddot{\mathcal{R}}_c+3H\dot{\mathcal{R}}_c-(c_s^2/a^2)\partial^2\mathcal{R}_c=0$.
This is a low-energy EFT for a massless mode with sound speed $c_s$. For $h\gg1$, $c_s^2\simeq -{1/3}$, which means $c_s$ is imaginary. The EFT applies after a mode is redshifted to $p\sim A|m_s|$, during which the curvature mode undergoes transient exponential growth, $\mathcal{R}\sim\exp(-k|c_s|\eta)$, before horizon crossing \cite{Chen:2023bcz}.

The cubic Lagrangian after integrating out heavy fields is given by
\begin{align}\label{L3fin}
    \mathcal{L}^{(3)}_{\text{eff}}={}\frac{a^3}{M_{\text{pl}}\sqrt{2\epsilon}}\bigg[\frac{\Lambda_1}{H}\dot{\mathcal{R}}_c^3+\frac{\Lambda_2}{a^2H}\dot{\mathcal{R}}_c(\partial_i\mathcal{R}_c)^2+\Lambda_3\mathcal{R}_c\dot{\mathcal{R}}_c^2-4\Xi_{\mathbb{U}}\left(3\mathcal{R}_c+\frac{1}{H}\dot{\mathcal{R}}_c\right)\bigg]+\mathcal{L}^{(3)}_{\text{bdy}},
\end{align}
where the temporal boundary terms are
\begin{align}
    \mathcal{L}^{(3)}_{\text{bdy}}=\frac{a^3}{M_{\text{pl}}\sqrt{2\epsilon}}\partial_t\left(\Lambda_4 H\mathcal{R}_c^3+\Lambda_5\mathcal{R}_c^2\dot{\mathcal{R}}_c\right),
\end{align}
$\Lambda_1\equiv 4$, $\Lambda_2\equiv2$, $\Lambda_3\equiv6(6-1/|c_s^2|)\simeq18$, $\Lambda_4\equiv108$, and $\Lambda_5\equiv3/|c_s^2|\simeq9$; spatial boundary terms have been omitted. The last term of the bulk interactions contains interactions with the magnetic perturbation $\mathbb{U}$. As shown in the next section, these interactions are subleading in the total non-Gaussianities. 

There are also corrections like higher-order of $\delta_{\square}$ and $1/h^2$ in the cubic action.
However we show in (\ref{L3firstorder}) that the coefficients of these correction operators are of the same order as those of the dominant operator. Thus, when the EFT becomes valid, namely at $p=p_i$, one has $\delta_{\square},1/h^2\ll1$, and the cubic Lagrangian (\ref{L3fin}) dominates the contribution. We also find that the $1/h^2$ correction terms contain some operators that are absent in the dominant terms, such as $\dot{\mathcal{R}}(\partial_i\mathcal{R})^2$. However, as we have seen above, the EFT is sufficiently valid only when $h\gtrsim5$, which is also the regime of interest to us. In this case 
$1/h^2\lesssim0.04$, which is small enough so that we can safely neglect these contributions.

So far we have derived the quadratic and cubic Lagrangians in both the small- and large-$h$ regimes. For small $h$, non-Gaussianities is generated by interactions between curvature and entropic perturbations and is organized as powers of $h$. For large $h$, the interaction coefficients in the EFT depend only weakly on $h$; the significant $h$ dependence instead enters through the mode evolution during the EFT regime. To see their dependence in non-Gaussianities, we now compute the bispectrum.

\section{Bispectrum}
We now calculate the non-Gaussianities of this model. Here we consider here the comoving curvature perturbation in the spatially-flat gauge, which is gauge-invariant for first order. However, we need to consider how the result is affected by nonlinear gauge transformations. So, we concern about the gauge-invariant curvature perturbation $\zeta$ in the uniform-density gauge up to second order. In other words,
\begin{align}\label{zeta3point}
    \langle\zeta(\boldsymbol{x})\zeta(\boldsymbol{y})\zeta(\boldsymbol{z})\rangle=-\langle\mathcal{R}(\boldsymbol{x})\mathcal{R}(\boldsymbol{y})\mathcal{R}(\boldsymbol{z})\rangle+\frac{1}{2}\langle\mathcal{R}(\boldsymbol{x})\mathcal{R}(\boldsymbol{y})\zeta^{(2)}(\boldsymbol{z})\rangle+2\rm perm.
\end{align}
where $\zeta=\zeta^{(1)}+\zeta^{(2)}/2+\cdots$ and we have used $\zeta^{(1)}\simeq-\mathcal{R}$ on large scales. We can calculate all cubic operators of $\mathcal{R}$ and then treat the non-linear term separately.

The expectation value of an operator $Q(\eta)$ is computed using the in-in formalism:
\begin{align}
    \langle \Omega | Q(\eta) | \Omega \rangle = \sum_{n=0}^\infty i^n &\int_{\eta_0}^{\eta} d\eta_1 \int_{\eta_0}^{\eta_1} d\eta_2 \cdots \int_{\eta_0}^{\eta_{n-1}} d\eta_n \nonumber\\
    &\times\left\langle [H_I(\eta_n), [H_I(\eta_{n-1}), \cdots, [H_I(\eta_1), Q_I(\eta)] \cdots ]] \right\rangle.
\end{align}
We consider the correlation function of the curvature perturbation,
$Q=\mathcal{R}_{\boldsymbol{k}_1}\mathcal{R}_{\boldsymbol{k}_2}\mathcal{R}_{\boldsymbol{k}_3}$, where $\mathcal{R}_{\boldsymbol{k}}$ is the Fourier transform of $\mathcal{R}$ and can be decomposed as $\mathcal{R}_{\boldsymbol{k}}=\mathcal{R}_{k}\hat{a}_{\boldsymbol{k}}+\mathcal{R}_{k}^{*}\hat{a}^{\dagger}_{-\boldsymbol{k}}$. The interaction Hamiltonian $H_I$ is provided by cubic Lagrangian (\ref{L_smallh}) for small $h$ and by (\ref{L3fin}) for large $h$. Then the bispectrum is defined as
\begin{align}
    \langle\mathcal{R}_{\boldsymbol{k}_1}\mathcal{R}_{\boldsymbol{k}_2}\mathcal{R}_{\boldsymbol{k}_3}\rangle\equiv(2\pi)^3\delta^{(3)}(\boldsymbol{k}_1+\boldsymbol{k}_2+\boldsymbol{k}_3)B_{\mathcal{R}}(k_1,k_2,k_3),
\end{align}
and one can discuss the shape function $S$, which is defined by the bispectrum as
\begin{align}
    B_{\mathcal{R}}(k_1,k_2,k_3)\equiv(2\pi)^4\frac{S(k_1,k_2,k_3)}{(k_1k_2k_3)^2}\mathcal{P}_{\mathcal{R}}^2.
\end{align}
where $\mathcal{P}_{\mathcal{R}}$ is the power spectrum of $\mathcal{R}$.

On the other hand, for non-linear contribution of $\zeta$, the uniform-density curvature perturbation up to second order is given by \cite{Funakoshi:2012ym},
\begin{align}
    \zeta^{(1)}&=-\frac{\mathcal{H}}{\rho_0'}\delta{\rho}^{(1)},\\
    \zeta^{(2)}&\simeq-\frac{\mathcal{H}}{\rho_{0}'}\delta{\rho}^{(2)}-2\frac{\delta\rho^{(1)}{'}}{\rho_0'}\zeta^{(1)}+2\zeta^{(1)}\zeta^{(1)},
\end{align}
where ``$\simeq$'' means we have ignored all slow-roll suppressed terms and gradient terms in the slow-roll limit and large-scale limit. Here $\rho_{0}$ is the background energy density and $\delta\rho$ is its perturbation. The terms on the r.h.s are written in the spatially-flat gauge. The second order $\delta\rho^{(2)}$ is given by 
\begin{align}
    \delta\rho^{(2)}\simeq\frac{3c^2}{2a^4f^2}\frac{(f^2)_{\phi\phi}}{f^2}\delta\phi^2+\frac{6c}{a^4}\frac{(f^2)_{\phi}}{f^2}\delta\phi\delta A'+\frac{3f^2}{a^4}(\delta A')^2,
\end{align}
where $c=\sqrt{2\epsilon_{\phi}}M_{\rm pl}h\mathcal{H}af$. Using $\rho_0'=-6\epsilon\mathcal{H}^3M_{\rm pl}^2/a^2$ and $\delta Q=\sqrt{2}f\delta A/a$, $\zeta^{(2)}$ can be reduced to
\begin{align}\label{zeta2}
    \zeta^{(2)}\simeq{}&\frac{1}{\epsilon M_{\rm pl}^2}\left(4h^2\delta\phi^2+6\sqrt{2}h\delta\phi\delta Q+\frac{9}{4}\delta Q^2\right)+2\zeta^{(1)}\zeta^{(1)}\nonumber\\
    &+\frac{1}{\epsilon M_{\rm pl}^2}\left(-2\sqrt{2}h\tau\delta\phi\delta Q'-\frac{3}{2}\tau\delta Q\delta Q'+\frac{1}{4}\tau^2\delta Q'\delta Q'\right)-2\eta\zeta^{(1)}{'}\zeta^{(1)}
\end{align}
Note that the bispectrum is computed on super-horizon scales hence all terms in the second line of (\ref{zeta2}) vanish because these perturbations are nearly constant on super-horizon scales. Writing down $\zeta^{(2)}$ in terms of the curvature perturbation $\mathcal{R}$ and entropy perturbation $\mathcal{F}$
\begin{align}
    \zeta^{(2)}\dot{\simeq}\frac{1}{\epsilon M_{\rm pl}^2}\left[-\frac{7h^2}{2(1+2h^2)}\mathcal{R}_c^2+\frac{\sqrt{2}h(3-8h^2)}{2(1+2h^2)}\mathcal{R}_c\mathcal{F}+\frac{8h^4+12h^2+9/4}{1+2h^2}\mathcal{F}^2\right]+2\mathcal{R}^2,
\end{align}
where ``$\dot{\simeq}$'' means we have discarded all time-derivative terms and we also use $\zeta^{(1)}\simeq-\mathcal{R}$ for large-scale modes.

To compute the bispectrum, we need to know the linear solutions of the perturbations. Solutions with real and imaginary sound speeds must be treated separately. For the general solution
\begin{align}
    \mathcal{R}_k(\eta)=\frac{A_k}{k^{3/2}}e^{-ikc_s\eta}\left(-ikc_s\eta-1\right)+\frac{B_k}{k^{3/2}}e^{ikc_s\eta}\left(ikc_s\eta-1\right).
\end{align}
the relation between the coefficients is fixed by the quantization condition
\begin{align}\label{qc}
    \mathcal{R}_k\mathcal{R}_k'^{*}-\mathcal{R}_k'\mathcal{R}_k^{*}=\frac{ic_s^2}{2\epsilon a^2M_{\text{pl}}^2},
\end{align}
where $'\equiv d/d\eta$. An imaginary sound speed fixes the coefficients at the initial time differently from a real one. For $h\ll1$, we compute vector-field corrections perturbatively around the free massless solution with $c_s=1$. For $h\gg1$, we instead use the single-mode EFT with imaginary sound speed. We discuss the contribution of each interaction term in these two regimes separately and assume $k_1\geq k_2\geq k_3$ below.

\subsection{Small $h$}

We treat the $h=0$ part as the free theory, so that the interaction Hamiltonian is given by (\ref{H_smallh}). Corrections at higher order in $h$ can be evaluated perturbatively when $h\ll1$. Since the free curvature perturbation is massless with unit sound speed, the quantization condition gives
\begin{align}
    |A_k|^2-|B_k|^2=\frac{H^2}{4\epsilon M_{\text{pl}}^2}.
\end{align}
Choosing $B_k=0$ selects the Bunch-Davies vacuum. The mode solution is then
\begin{align}\label{mode_smallh}
    \mathcal{R}_k=\frac{H}{\sqrt{2k^{3}}}e^{-ik\eta}\left(ik\eta+1\right),
\end{align}
and the power spectrum is given by
\begin{align}\label{PS_smallh}
    \mathcal{P}_{\mathcal{R}}=\frac{H^2}{8\pi^2\epsilon M_{\text{pl}}^2}.
\end{align}

\begin{figure}[t]
\centering
\includegraphics[scale=0.7]{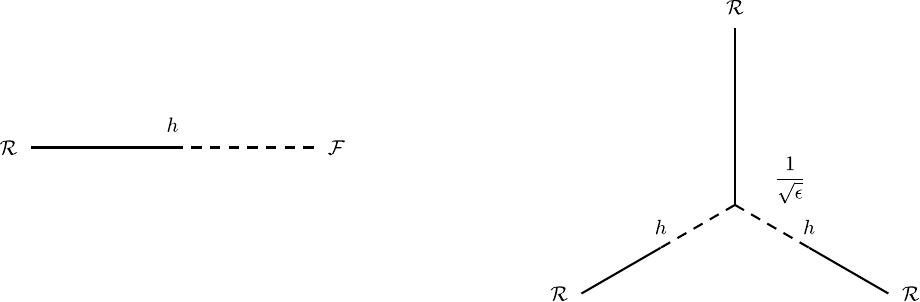}
\caption{\label{fig:vertex}(left) Vertex of exchanging $\mathcal{R}$ and $\mathcal{F}$. (right) The dominated diagram contributing to the non-Gaussianities for $h\ll1$.}
\end{figure}

The quadratic interaction describes the exchange vertex between $\mathcal{R}_c$ and $\mathcal{F}$ shown in FIG.~\ref{fig:vertex}(left), whose coupling strength is $\sim h$. The leading non-Gaussian contribution is provided by the diagram in FIG.~\ref{fig:vertex}(right) \cite{Gorji:2020vnh,Funakoshi:2012ym}. Accordingly, the relevant interaction Hamiltonian is $H_I=H_I^{(2)}+H_I^{(3)}$, where
\begin{align}
    H_I^{(2)}\doteq{}&\int d^3x\ 4\sqrt{2}a^3hH\mathcal{F}\dot{\mathcal{R}}_c,\\
    H_I^{(3)}\doteq{}&\int d^3x\ \frac{a^3}{M_{\text{pl}}\sqrt{2\epsilon}}\left(\frac{4}{3}\mathcal R_c\dot{\mathcal F}^{2}+12H^2\mathcal R_c\mathcal F^2\right),
\end{align}
and ``$\doteq$'' indicates that only vertices contributing to FIG.~\ref{fig:vertex} are retained. The corresponding three-point function is
\begin{align}
   \langle\mathcal{R}_{\boldsymbol{k}_1}\mathcal{R}_{\boldsymbol{k}_2}\mathcal{R}_{\boldsymbol{k}_3}\rangle\supset{}& -i \int_{\eta_0}^{\eta_e} d\eta_1 \int_{\eta_0}^{\eta_1} d\eta_2 \int_{\eta_0}^{\eta_2} d\eta_3\nonumber\\
   &\times\Bigg\{
\Big\langle \Big[  H_{I}^{(3)}(\eta_3), \Big[H_{I}^{(2)}(\eta_2), \Big[H_{I}^{(2)}(\eta_1), \mathcal{R}_{\boldsymbol{k}_1}(\eta_e)\mathcal{R}_{\boldsymbol{k}_2}(\eta_e)\mathcal{R}_{\boldsymbol{k}_3}(\eta_e) \Big] \Big] \Big] \Big\rangle\nonumber
\\
&+ \Big\langle \Big[H_{I}^{(2)}(\eta_3), \Big[H_{I}^{(3)}(\eta_2), \Big[ H_{I}^{(2)}(\eta_1), \mathcal{R}_{\boldsymbol{k}_1}(\eta_e)\mathcal{R}_{\boldsymbol{k}_2}(\eta_e)\mathcal{R}_{\boldsymbol{k}_3}(\eta_e) \Big] \Big] \Big] \Big\rangle
\nonumber\\
&+ \Big\langle \Big[ H_{I}^{(2)}(\eta_3), \Big[H_{I}^{(2)}(\eta_2) , \Big[ H_{I}^{(3)}(\eta_1), \mathcal{R}_{\boldsymbol{k}_1}(\eta_e)\mathcal{R}_{\boldsymbol{k}_2}(\eta_e)\mathcal{R}_{\boldsymbol{k}_3}(\eta_e) \Big] \Big] \Big] \Big\rangle
\Bigg\}.
\end{align}
Using the mode function (\ref{mode_smallh}) and power spectrum (\ref{PS_smallh}), we obtain
\begin{align}
     \langle\mathcal{R}_{\boldsymbol{k}_1}\mathcal{R}_{\boldsymbol{k}_2}\mathcal{R}_{\boldsymbol{k}_3}\rangle\supset{}-256h^2N_K^3(2\pi^2)^2\mathcal{P}_{\mathcal{R}}^2(2\pi)^3\delta^{(3)}(\boldsymbol{k}_1+\boldsymbol{k}_2+\boldsymbol{k}_3)\frac{k_1^3+k_2^3+k_3^3}{k_1^3k_2^3k_3^3}+\mathcal{O}(N_K^2),
\end{align}
where $N_K\equiv-\log(-K\eta_e)$ and $K\equiv(k_1+k_2+k_3)/3$. The leading contribution arises from the second vertex of $H^{(3)}_I$. The first vertex gives a term proportional to $N_K^2$: the enhancement originates from the two exchange vertices, whereas the cubic vertex remains finite. In the local limit $k_1=k_2=k$, $k_3=q$, with $r\equiv q/k\ll1$, the shape function becomes
\begin{align}
    S_{h\ll1}(k,k,q)=-\frac{128h^2N_K^3}{r}+\mathcal{O}(r^2),
\end{align}
which diverges at $r=0$. In other words, the bispectrum is of local shape.

This result shows that vector fields source the curvature perturbation cumulatively on superhorizon scales: $\mathcal{R}$ continues to evolve after horizon crossing. For $N_K\sim\mathcal{O}(100)$, CMB bounds consequently impose the stringent constraint $h\lesssim\mathcal{O}(10^{-4})$. In the large-$h$ regime, however, the cumulative mode decays and is replaced by a constant mode \cite{Chen:2023bcz}. We next calculate the bispectrum of these constant modes.

\subsection{Large $h$}
In the EFT regime, the dominant contribution comes from a one-vertex diagram. The leading three-point correlation function of $\mathcal{R}$ is therefore
\begin{align}\label{3point_largeh}
    \langle\mathcal{R}_{\boldsymbol{k}_1}\mathcal{R}_{\boldsymbol{k}_2}\mathcal{R}_{\boldsymbol{k}_3}\rangle\simeq{}2\text{Im}\left[\int_{\eta_i}^{\eta_e}d\eta'a(\eta')\langle\mathcal{R}_{\boldsymbol{k}_1}(\eta_e)\mathcal{R}_{\boldsymbol{k}_2}(\eta_e)\mathcal{R}_{\boldsymbol{k}_3}(\eta_e)H^{(3)}_{I,\ \text{eff}}(\eta')\rangle\right]
\end{align}
where $\eta_i\equiv-x/(c_sk_m)$, $\eta_e\to0$, $k_m=\max(k_1,k_2,k_3)$ and 
\begin{align}
    H^{(3)}_{I,\ \text{eff}}=-\int d^3x\mathcal{L}^{(3)}_{\text{eff}}.
\end{align}
The lower limit $\eta_i$ is imposed because the EFT becomes valid only after the physical momentum has redshifted to $p_i/|m_s|=xH/(|c_s||m_s|)=A$. The e-folds number between the validity of EFT and the horizon-crossing is given by $\Delta N=\ln x$. By substituting the sound speed and the entropy mass, the relation between $x$ and $h$ can be derived as
\begin{align}
    x\simeq A\sqrt{\frac{2}{3}}\left(2+\frac{7}{3h^2}\right)h+\mathcal{O}(\frac{1}{h^3}).
\end{align}
For $h\gtrsim5$, where the EFT is valid, the $1/h^2$ and higher-order corrections are negligible. In this regime, the reliability of the EFT is dictated by the magnitude of the higher-derivative correction $\delta_{\square}$, which is essentially set by $A^2$. As an example, taking $A=0.3$ yields $x\simeq0.5h$. In this section, we restrict our calculation to the leading-order contributions. An estimate of the higher-derivative operators to solution (\ref{F_corr}) in the cubic action is relegated to the appendix \ref{appendix2}, as these corrections are subleading and are not expected to appreciably modify the bispectrum structure.

For $h\gg1$, the sound speed satisfies $c_s^2\simeq-1/3$ and is therefore imaginary. Quantization for a purely imaginary $c_s$ differs from the real-$c_s$ case. Imposing (\ref{qc}) gives
\begin{align}
    \text{Im}\left[A_k^*B_k\right]=\frac{H^2}{8\epsilon|c_s|M{\text{pl}}^2}.
\end{align}
The mode solution can be parametrized as \cite{Garcia-Saenz:2018ifx}
\begin{align}
    &\mathcal{R}_k=\frac{\alpha_k}{k^{3/2}}\Big[e^{k|c_s|\eta+x}(k|c_s|\eta-1)-e^{\gamma_k+i\theta_k}e^{-(k|c_s|\eta+x)}(k|c_s|x\eta+1)\Big],
\end{align}
where all parameters are real and satisfy the quantization condition
\begin{align}\label{QC}
    \alpha_k^2e^{\gamma_k}\sin(\theta_k)=\frac{H^2}{8\epsilon|c_s|M_{\text{pl}}^2}.
\end{align}
The EFT becomes valid at $k|c_s\eta|=x$ for a mode of wavenumber $k$. For $h\gg1$, the power spectrum of $\mathcal{R}$ is
\begin{align}\label{PS}
    \mathcal{P}_{\mathcal{R}}\simeq \frac{\alpha_k^2}{2\pi^2}e^{2x}.
\end{align}

Each $\mathcal{R}$ in the integral contains one growing factor $e^x$ and one decaying factor $e^{-x}$. Since $x\sim h\gg1$, the nominally leading term scales as $e^{6x}$, but it is real and does not contribute to the final correlator. The leading non-vanishing term is therefore of order $e^{4x}$ and contains one decaying-mode insertion. The second line of (\ref{L3fin}) describes interactions with the magnetic perturbation. Since $\mathbb{U}$ decouples at quadratic order and remains massless, it is not exponentially amplified near horizon crossing. Its contributions are at most of order $e^{3x}$ and are neglected below.

\textbf{Interaction in $\dot{\mathcal{R}}^3$: } We first evaluate the $\dot{\mathcal{R}}^3$ term in the interaction Hamiltonian. Using (\ref{3point_largeh}), we obtain
\begin{align}\label{Sdot3}
    \langle\mathcal{R}_{\boldsymbol{k}_1}\mathcal{R}_{\boldsymbol{k}_2}\mathcal{R}_{\boldsymbol{k}_3}\rangle
    \supset{}
    ={}&24\Lambda_1\epsilon M_{\text{pl}}^2(2\pi)^3\delta^{(3)}(\boldsymbol{k}_1+\boldsymbol{k}_2+\boldsymbol{k}_3)\nonumber\\
    &\times\text{Im}\left[\int^{\eta_e}_{\eta_i}\frac{d\eta'}{\eta'}\mathcal{R}_{k_1}(\eta_e)\mathcal{R}_{k_2}(\eta_e)\mathcal{R}_{k_3}(\eta_e)\mathcal{R}_{k_1}^{*'}(\eta')\mathcal{R}_{k_2}^{*'}(\eta')\mathcal{R}_{k_3}^{*'}(\eta')\right]\nonumber\\
    \simeq{}&6\Lambda_1|c_s|^2(2\pi^2)^2\mathcal{P}_{\mathcal{R}}^2(2\pi)^3\delta^{(3)}(\boldsymbol{k}_1+\boldsymbol{k}_2+\boldsymbol{k}_3)\frac{1}{(k_1k_2k_3)^2}\nonumber\\
    &\times\Bigg\{-\frac{k_1k_2k_3}{(k_1+k_2+k_3)^3}
    +\frac{k_1k_2k_3}{\tilde{k}_1^3}\bigg[1-e^{-\tilde{k}_1x/k_m}\bigg(1+\frac{\tilde{k}_1x}{k_m}+\frac{\tilde{k}_1^2x^2}{2k_m^2}\bigg)\bigg]\Bigg\}\nonumber\\
    &+\text{2 perm.},
\end{align}
where terms proportional to $e^{-(k_1+k_2+k_3)x/k_m}$ have been neglected and (\ref{QC}) and (\ref{PS}) have been used. Here $\tilde{k}_1=-k_1+k_2+k_3$. A contribution of this form was first obtained in multi-field inflation with a non-geodesic trajectory in field space \cite{Garcia-Saenz:2019njm,Garcia-Saenz:2018ifx,Fumagalli:2019noh}. Its exponential dependence, $e^{-\tilde{k}_ix/k_m}$, is a consequence of the imaginary sound speed and becomes important in the flattened limit. In particular,
\begin{align}
    S_{\dot{\mathcal{R}}^3}(k_2+k_3,k_2,k_3)\simeq\frac{\Lambda_1|c_s|^2}{4}\frac{k_2k_3}{k_1^2}x^3,
\end{align}
which is enhanced by $x^3$. This enhancement arises from inserting one decaying mode in the internal leg with momentum $k_1$, producing the factor $e^{-\tilde{k}_1x/k_m}$. The other two terms, $e^{-\tilde{k}_2x/k_m}$ and $e^{-\tilde{k}_3x/k_m}$, are of order $\mathcal{O}(e^{-x})$ and have been discarded. The first term in (\ref{Sdot3}), generated by a decaying-mode insertion on an external leg, is of order $\mathcal{O}(1)$. The shape $S_{\dot{\mathcal{R}}^3}$ is finite in the equilateral limit and vanishes in the local limit; for large $x$, it is dominated by flattened configurations.

\textbf{Interaction in $\dot{\mathcal{R}}(\partial_i\mathcal{R})^2$: } We next compute the $\dot{\mathcal{R}}\partial_i\mathcal{R}\partial_i\mathcal{R}$ term in the interaction Hamiltonian. Using (\ref{3point_largeh}), we obtain
\begin{align}
    \langle\mathcal{R}_{\boldsymbol{k}_1}\mathcal{R}_{\boldsymbol{k}_2}\mathcal{R}_{\boldsymbol{k}_3}\rangle
    \supset{}
    &-\frac{8\Lambda_2\epsilon M_{\text{pl}}^2}{H^2}(2\pi)^3\delta^{(3)}(\boldsymbol{k}_1+\boldsymbol{k}_2+\boldsymbol{k}_3)(\boldsymbol{k}_2\cdot\boldsymbol{k}_3)\nonumber\\
    &\times\text{Im}\bigg[\int_{\eta_i}^{\eta_e}\frac{d\eta'}{\eta'}\mathcal{R}_{k_1}(\eta_e)\mathcal{R}_{k_2}(\eta_e)\mathcal{R}_{k_3}(\eta_e)\mathcal{R}_{k_1}^{*'}(\eta')\mathcal{R}_{k_2}^{*}(\eta')\mathcal{R}_{k_3}^{*}(\eta')\bigg]+\text{2 perm.}\nonumber\\
    \simeq&-\frac{1}{2}\Lambda_2(2\pi^2)^2\mathcal{P}_{\mathcal{R}}^2(2\pi)^3\delta^{(3)}(\boldsymbol{k}_1+\boldsymbol{k}_2+\boldsymbol{k}_3)\frac{1}{(k_1k_2k_3)^2}\nonumber\\
    &\times\Bigg\{-\frac{k_1k_2k_3}{(k_1+k_2+k_3)^3}p_0(k_1,k_2,k_3)+\frac{k_1k_2k_3}{\tilde{k}_1^3}\bigg[p_0(-k_1,k_2,k_3)\nonumber\\
    &-e^{-\tilde{k}_1x/k_m}\bigg(p_0(-k_1,k_2,k_3)+p_1(-k_1,k_2,k_3)\frac{\tilde{k}_1x}{k_m}+p_2(-k_1,k_2,k_3)\frac{\tilde{k}_1^2x^2}{2k_m^2}\bigg)\bigg]\Bigg\}\nonumber\\
    &+\text{2 perm.},
\end{align}
where
\begin{align}
    p_0(k_1,k_2,k_3)={}&12+9\bigg(\frac{k_1}{k_2}+\text{5 perm.}\bigg)+\bigg(\frac{k_1^2}{k_2^2}+\text{5 perm.}\bigg)-6\bigg(\frac{k_1^2}{k_2k_3}+\text{2 perm.}\bigg)\nonumber\\
    &+6\bigg(\frac{k_1k_2}{k_3^2}+\text{2 perm.}\bigg)-3\bigg(\frac{k_1^3}{k_2k_3^2}+\text{5 perm.}\bigg)-\bigg(\frac{k_1^4}{k_2^2k_3^2}+\text{2 perm.}\bigg),\nonumber\\
    p_1(k_1,k_2,k_3)={}&6+5\bigg(\frac{k_1}{k_2}+\text{5 perm.}\bigg)-4\bigg(\frac{k_1^2}{k_2^2}+\text{5 perm.}\bigg)+2\bigg(\frac{k_1k_2}{k_3^2}+\text{2 perm.}\bigg)\nonumber\\
    &-\bigg(\frac{k_1^3}{k_2k_3^2}+\text{5 perm.}\bigg),\nonumber\\
    p_2(k_1,k_2,k_3)={}&2\bigg(\frac{k_1}{k_2}+\text{5 perm.}\bigg)-2\bigg(\frac{k_1^2}{k_2^2}+\text{5 perm.}\bigg).
\end{align}
We used $2\boldsymbol{k}_1\cdot\boldsymbol{k}_2=k_2^2+k_3^2-k_1^2$ in the calculation. This shape is likewise enhanced by a decaying-mode insertion on an internal leg. For $k_1=k_2+k_3$,
\begin{align}
    S_{\dot{\mathcal{R}}(\partial_i\mathcal{R})^2}(k_2+k_3,k_2,k_3)\simeq \frac{\Lambda_2}{4}\frac{k_2k_3}{k_1}x^3.
\end{align}
The other two permutations are of order $\mathcal{O}(e^{-x})$ in this limit. Its equilateral and local limits are negligible relative to its flattened limit. This contribution also occurs in multi-field inflation with a non-geodesic trajectory \cite{Garcia-Saenz:2019njm,Garcia-Saenz:2018ifx,Fumagalli:2019noh} and is therefore a familiar flattened signal. The vector-field model additionally produces the interaction $\mathcal{R}\dot{\mathcal{R}}^2$ and a temporal boundary term, to which we now turn.

\textbf{Interaction in $\mathcal{R}\dot{\mathcal{R}}^2$: } We compute the $\mathcal{R}\dot{\mathcal{R}}^2$ term in the interaction Hamiltonian using (\ref{3point_largeh}):
\begin{align}
    \langle\mathcal{R}_{\boldsymbol{k}_1}\mathcal{R}_{\boldsymbol{k}_2}\mathcal{R}_{\boldsymbol{k}_3}\rangle
    \supset{}
    {}&-\frac{8\Lambda_3\epsilon M_{\text{pl}}^2}{H^2}(2\pi)^3\delta^{(3)}(\boldsymbol{k}_1+\boldsymbol{k}_2+\boldsymbol{k}_3)\nonumber\\
    &\times\text{Im}\bigg[\int_{\eta_i}^{\eta_e}\frac{d\eta'}{\eta'^2}\mathcal{R}_{k_1}(\eta_e)\mathcal{R}_{k_2}(\eta_e)\mathcal{R}_{k_3}(\eta_e)\mathcal{R}_{k_1}^{*}(\eta')\mathcal{R}_{k_2}^{*'}(\eta')\mathcal{R}_{k_3}^{*'}(\eta')\bigg]+\text{2 perm.}\nonumber\\
    \simeq{}&-\Lambda_3|c_s|^2(2\pi^2)^2\mathcal{P}_{\mathcal{R}}^2(2\pi)^3\delta^{(3)}(\boldsymbol{k}_1+\boldsymbol{k}_2+\boldsymbol{k}_3)\frac{1}{(k_1k_2k_3)^2}\nonumber\\
    &\Bigg\{\frac{k_1k_2k_3}{(k_1+k_2+k_3)^3}q_0(k_1,k_2,k_3)-\frac{k_1k_2k_3}{\tilde{k}_1^3}\bigg[q_0(-k_1,k_2,k_3)\nonumber\\
    &-e^{-\tilde{k}_1x/k_m}\bigg(q_0(-k_1,k_2,k_3)+q_1(-k_1,k_2,k_3)\frac{\tilde{k}_1x}{k_m}\bigg)\bigg]\Bigg\}\nonumber\\
    &+\text{2 perm.},
\end{align}
where
\begin{align}
    q_0(k_1,k_2,k_3)={}&6+3\bigg(\frac{k_1}{k_2}+\text{5 perm.}\bigg)+\bigg(\frac{k_1^2}{k_2^2}+\text{5 perm.}\bigg)+2\bigg(\frac{k_1k_2}{k_3^2}+\text{2 perm.}\bigg),\nonumber\\
    q_1(k_1,k_2,k_3)={}&3+\bigg(\frac{k_1}{k_2}+\text{5 perm.}\bigg).
\end{align}
This shape is also enhanced in the flattened limit through a decaying-mode insertion. For $\tilde{k}_1=0$, we obtain
\begin{align}
    S_{\mathcal{R}\dot{\mathcal{R}}^2}(k_2+k_3,k_2,k_3)\simeq\frac{\Lambda_3|c_s|^2}{4}\frac{k_2^2+k_2k_3+k_3^2}{2k_1^2}x^2,
\end{align}
which is enhanced by $\sim x^2$. The contribution from a decaying-mode insertion on an external leg is of order $\mathcal{O}(1)$, whereas those associated with the internal legs $k_2$ and $k_3$ are of order $\mathcal{O}(e^{-x})$ and have been neglected.

On the other hand, the equilateral limit of this shape is 
\begin{align}
    S_{\mathcal{R}\dot{\mathcal{R}}^2}(k, k, k)\simeq \Lambda_3|c_s|^2\left[2-\frac{3}{4}\left(4+x\right)e^{-x}\right],
\end{align}
which is finite for large $x$.

Unlike the first two interactions, this one contains a local component. Consider $k_1=k_2=k$, $k_3=q$, and define $r\equiv q/k\ll1$. The first term, from a decaying-mode insertion on an external leg, contributes $\sim1/r$. Insertions giving $k_1\to-k_1$ and $k_2\to-k_2$ contribute $\sim x/r$, while the $k_3\to-k_3$ insertion contributes $\sim1/r$. Combining these terms yields
\begin{align}
    S_{\mathcal{R}\dot{\mathcal{R}}^2}(k,k,q)\simeq \frac{\Lambda_3|c_s|^2}{8}\left(4x-2-e^{-2x}\right)\frac{1}{r}+\mathcal{O}(r),
\end{align}
which diverges as $r\to0$. Since the third term $e^{-2x}$ is suppressed for large $x$, the amplitude of this local component grows linearly with $x$. We emphasize that this local limit gives the EFT bulk contribution
to the squeezed coefficient for the chosen matching prescription. However, this coefficient is not complete; we also need to consider the additional effects arising from different momenta entering the EFT regime at different times and metric perturbations. We will briefly discuss these issues at the end of this section. Our conclusion is that these effects do not alter the $x$-enhancement feature of the local shape.

\textbf{Interaction in $\mathcal{H}^{(3)}_{\text{bdy}}$: } 
We finally compute the boundary term. Its first term commutes with $\mathcal{R}_{\boldsymbol{k}}$ and does not contribute to the three-point correlation function. The second term gives, using (\ref{3point_largeh}),
\begin{align}
    \langle\mathcal{R}_{\boldsymbol{k}_1}\mathcal{R}_{\boldsymbol{k}_2}\mathcal{R}_{\boldsymbol{k}_3}\rangle
    \supset{}
    {}&-8\Lambda_{5}\epsilon M_{\text{pl}}^2(2\pi)^3\delta^{(3)}(\boldsymbol{k}_1+\boldsymbol{k}_2+\boldsymbol{k}_3)\nonumber\\
    &\times\text{Im}\bigg[\mathcal{R}_{\boldsymbol{k}_1}(\eta_e)\mathcal{R}_{\boldsymbol{k}_2}(\eta_e)\mathcal{R}_{\boldsymbol{k}_3}(\eta_e)\mathcal{R}^{*'}_{k_1}(\eta')\mathcal{R}^{*}_{k_2}(\eta')\mathcal{R}^{*}_{k_3}(\eta')\bigg]\Bigg|_{\eta_i}^{\eta_e}+\text{2 perm.}\nonumber\\
    \simeq{}&\Lambda_5|c_s|^2(2\pi^2)^2\mathcal{P}_{\mathcal{R}}^2(2\pi)^3\delta^{(3)}(\boldsymbol{k}_1+\boldsymbol{k}_2+\boldsymbol{k}_3)\frac{1}{(k_1k_2k_3)^2}\Bigg[2\bigg(\frac{k_1^2}{k_2k_3}\bigg)\nonumber\\
    &-e^{-\tilde{k}_1x/k_m}\bigg(-\frac{k_m}{\tilde{k}_1x}l_{-1}(-k_1,k_2,k_3)-l_0(-k_1,k_2,k_3)-\frac{\tilde{k}_1x}{k_m}\bigg)\Bigg]\nonumber\\
    &+\text{2 perm.},
\end{align}
where
\begin{align}
    l_{-1}(k_1,k_2,k_3)=&\bigg(\frac{k_1}{k_2}+\text{5 perm.}\bigg)+\bigg(\frac{k_1^2}{k_2k_3}+\text{2 perm.}\bigg),\nonumber\\
    l_{0}(k_1,k_2,k_3)=&\frac{k_1}{k_2}+\text{5 perm.}
\end{align}
This boundary contribution is not enhanced in the flattened limit. For $\tilde{k}_1=0$, we have
\begin{align}
    S_{\text{bdy}}(k_2+k_3,k_2,k_3)\simeq\frac{\Lambda_5|c_s|^2}{4}\left[-2\left(\frac{k_1^2}{k_2k_3}+2 \text{ perm.}\right)+3+\frac{2}{x}\frac{k_2^2+k_2k_3+k_3^2}{k_2k_3}\right],
\end{align}
which becomes independent of $x$ at large $x$. The first contribution comes from a decaying-mode insertion on an external leg, while the remaining two contributions arise from the internal-leg insertion $k_1\to-k_1$. The other two permutations have been neglected because they are of order $\mathcal{O}(e^{-x})$.

The equilateral limit of this shape is given by
\begin{align}
    S_{\text{bdy}}(k, k, k)\simeq \frac{3}{2}-\frac{3}{4}\left(2-x+\frac{3}{x}\right)e^{-x},
\end{align}
which is also finite for large $x$.

As for the $\mathcal{R}\dot{\mathcal{R}}^2$ interaction, the boundary contribution also contains a non-negligible local component:
\begin{align}
    S_{\text{bdy}}(k,k,q)\simeq \frac{\Lambda_5|c_s|^2}{2}\left(2-\frac{2}{x}\right)\frac{1}{r}+\mathcal{O}(r).
\end{align}
The term $\sim1/r$ comes from internal-leg insertions with $k_1\to-k_1$ and $k_2\to-k_2$, while the term $\sim1/(xr)$ arises from an external-leg insertion. All contributions suppressed by $e^{-x}$ have been omitted. Unlike the bulk derivative interactions, this local amplitude is not enhanced by $x$ and remains finite at large $x$.

\textbf{Non-linear contribution: }For large-$h$, substituting the leading-order solution for $\mathcal{F}$, and dropping the time-derivative terms, this non-linear perturbation is reduced to
\begin{align}
    \zeta^{(2)}_{\rm LO}\simeq\Lambda_6\mathcal{R}^2,\ \ \ \ \ \ \ \Lambda_6\equiv2+\frac{9h^2(1+2h^2)}{(3+2h^2)^2},
\end{align}
Contributions to the correlation function become
\begin{align}
    \langle\zeta(\boldsymbol{x})\zeta(\boldsymbol{y})\zeta(\boldsymbol{z})\rangle\simeq\Lambda_6\langle\mathcal{R}(\boldsymbol{x})\mathcal{R}(\boldsymbol{z})\rangle \langle\mathcal{R}(\boldsymbol{y})\mathcal{R}(\boldsymbol{z})\rangle+2\rm\ perm.
\end{align}
and the corresponding bispectrum is
\begin{align}
    B_{\rm nl}(k_1, k_2, k_3)\simeq\Lambda_6(2\pi^2)^2\mathcal{P}_{\mathcal{R}}^2\left(\frac{1}{k_1^3k_2^3}+2\rm\ perm.\right).
\end{align}
In the local limit, the contribution due the non-linear transformation to shape functions is
\begin{align}
    S_{\rm nl}(k,k,q)\simeq\frac{\Lambda_6}{2r}.
\end{align}
It can be seen that for $h\gg1$, $\Lambda_6\simeq13/2$, so the contribution of this term is not enhanced by $x$. However, its contribution is still not negligible, and we therefore take it into account as well.

Before we end this section, we need to emphasize some remarks. For $r\ll1$, when the $k_L$ mode crosses the horizon, the $k_S$ mode is still deeply inside the UV region in which the short modes are rapidly oscillating. Hence, contributions from this region are suppressed by oscillatory cancellation and reduce mainly to matching-surface terms. The matching contribution is localized near the UV-to-EFT transition and does not acquire the cumulative enhancement generated by the subsequent bulk EFT evolution. Hence, we expect the omitted transition-region contribution to generate an $\mathcal{O}(1/r)$ matching term to shape functions without the cumulative $x$-enhancement of the EFT bulk interaction, i.e. $\Delta S=C_{\rm UV}/r$ with $C_{\rm UV}\sim\mathcal{O}(1)$. Here we have simply taken the crude limit $r\to0$, which is not exact. However, this does not affect the $x$-enhancement of the local shape, so we will not address this correction in the present work. To more accurately compute the local-shape contribution, we need a complete treatment of the UV theory. For the coupled system discussed in this work, knowledge of its analytical solution would help us address this issue \cite{Huenupi:2026abj}.

We also emphasize that we have ignored the metric perturbations in our calculations. The $x/r$ term obtained above represents the bulk matter-sector contribution within the low-energy EFT. A complete test of the consistency relation requires the simultaneous inclusion of the UV matching, initial-state contributions, and the interactions induced by the lapse and shift constraints. Although the interaction vertices induced by the lapse and shift constraints are slow-roll suppressed at the level of the action, their net contribution cannot be discarded a priori in a strict consistency-relation check. However, if one expects these contributions to be at most of order $\mathcal{O}(1)$ or $\mathcal{O}(\epsilon,\eta)$, so they will not cancel the $x/r$ term. An uncancelled enhanced squeezed contribution would indicate that at least one of the assumptions of the consistency conditions is not satisfied. We therefore regard the local momentum scaling and the bulk $x$-enhancement as results of the present EFT calculation, while the exact coefficient in the strict squeezed limit remains matching dependent. We leave a full discussion and check for future work.

\section{Non-Gaussianities for large $h$}
Having computed the interactions contributing to the bispectrum, we now examine the resulting shapes and amplitudes. We compare our shape functions with the standard equilateral, orthogonal, flattened, and local templates used in data analysis:
\begin{align}
    &S^{\text{eq}}=\frac{9}{10}\frac{\tilde{k}_1\tilde{k}_2\tilde{k}_3}{k_1k_2k_3},\ \ \ \ \ \ S^{\text{orth}}=3S^{\text{eq}}-\frac{9}{5},\ \ \ \ \ \ S^{\text{flat}}=-S^{\text{eq}}+\frac{9}{10},\nonumber\\
    &S^{\text{local}}=\frac{3}{10}\left(\frac{k_1^2}{k_2k_3}+2\text{ perm.}\right).
\end{align}

\begin{figure}[tbp]
  \centering
  \begin{minipage}{0.8\textwidth}
    \centering
    \includegraphics[width=\linewidth]{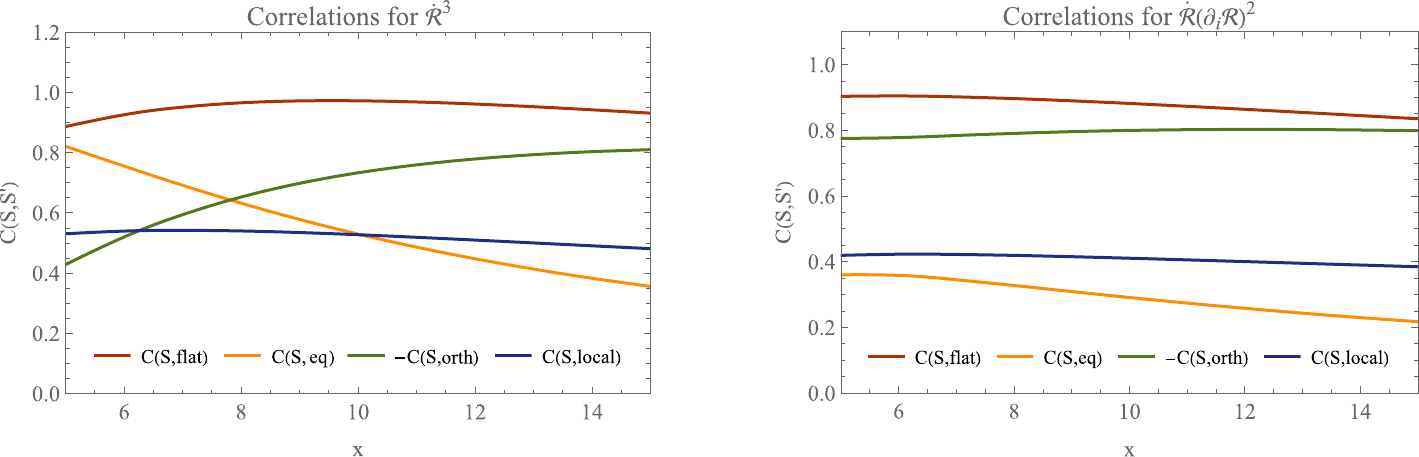}
    \caption{Correlations of the $\dot{\mathcal{R}}^3$ (left) and $\dot{\mathcal{R}}(\partial_i\mathcal{R})^2$ (right) contributions with standard nonlocal templates as functions of $x$.}
    \label{fig:Ceq}
  \end{minipage}

  \vspace{1em}  

  \begin{minipage}{0.8\textwidth}
    \centering
    \includegraphics[width=\linewidth]{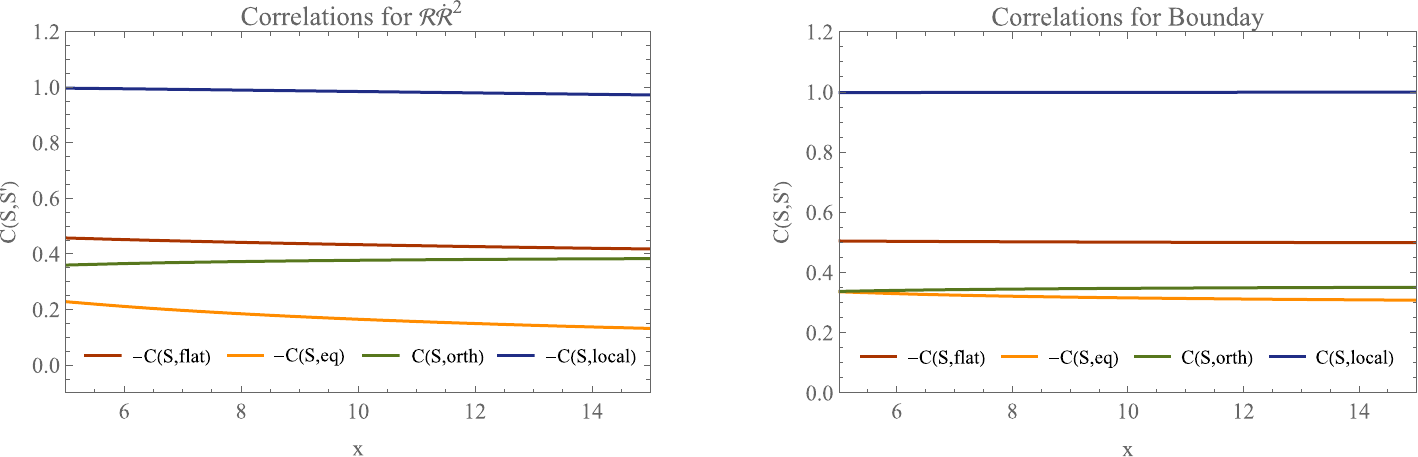}
    \caption{Correlations of the $\mathcal{R}\dot{\mathcal{R}}^2$ and boundary contributions with the local template and the remaining standard templates as functions of $x$.}
    \label{fig:Clocal}
  \end{minipage}
\end{figure}
\begin{figure}[tbp]
  \centering
  \begin{minipage}{0.8\textwidth}
    \centering
    \includegraphics[width=\linewidth]{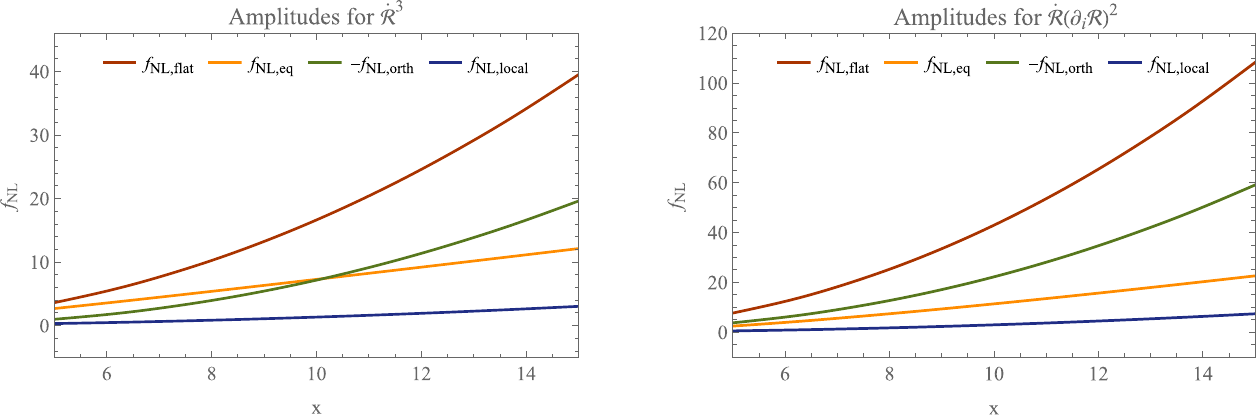}
    \caption{Projected amplitudes of the flattened-enhanced bulk interactions, shown without the coefficients $\Lambda_i$.}
    \label{fig:Aflat}
  \end{minipage}

  \vspace{1em} 

  \begin{minipage}{0.8\textwidth}
    \centering
    \includegraphics[width=\linewidth]{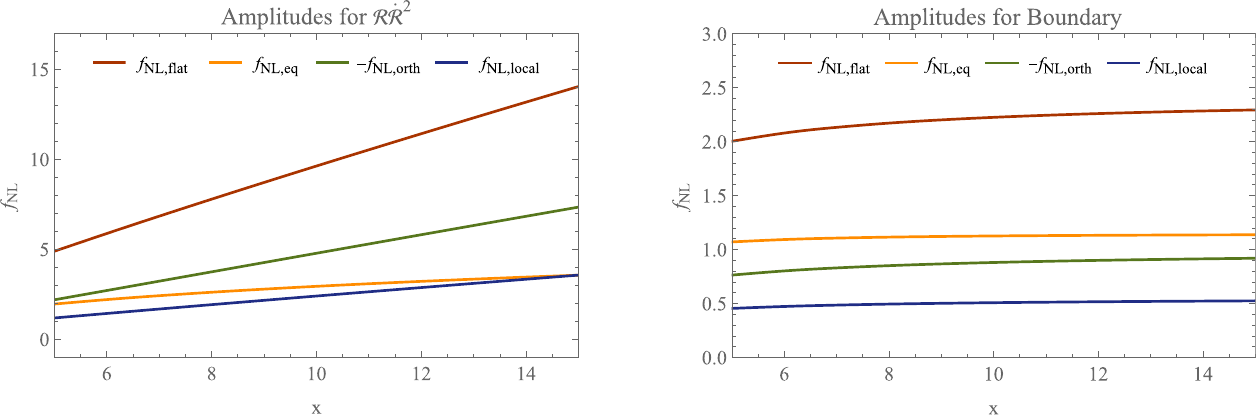}
    \caption{Projected amplitudes of the $\mathcal{R}\dot{\mathcal{R}}^2$ and boundary contributions, shown without the coefficients $\Lambda_i$.}
    \label{fig:Alocal}
  \end{minipage}
\end{figure}

We use the correlation function proposed in \cite{Babich:2004gb,Fergusson:2008ra} to quantify the similarity between two shapes. It is defined from the inner product
\begin{align}
    F(S, S') = \int_{\mathcal{V}_k} S(k_1, k_2, k_3) \, S'(k_1, k_2, k_3) \, \omega(k_1, k_2, k_3) \, d\mathcal{V}_k,
\end{align}
where the integration domain $\mathcal{V}_k$ is the set of all allowed momentum triangles: $\mathcal{V}_k=\{0\leq k_i\leq k_m,\ k_i\leq k_j+k_k\}$ due to the Dirac delta function $\delta^{(3)}(\boldsymbol{k}_1+\boldsymbol{k}_2+\boldsymbol{k}_3)$ in the bispectrum. The weight function $\omega$ is chosen as
\begin{align}
    \omega(k_1, k_2, k_3)=\frac{1}{k_1+k_2+k_3}.
\end{align}
Then the correlation function and the amplitude are given by 
\begin{align}
   C(S, S')=\frac{F(S, S')}{\sqrt{F(S, S)F(S', S')}},\ \ \ \ \ \ \ \ f_{\text{NL}}^{S}=\frac{F(S, S')}{F(S', S')}
\end{align}
respectively. We display the correlation for each interaction discussed in the previous section. We restrict attention to $x\sim h\gtrsim5$, where the EFT description is well behaved. The figures include only leading-order contributions; terms suppressed by $e^{-x}$ are discarded.

For the local shape, the correlation and amplitude integrals diverge when one momentum approaches zero. We therefore introduce an IR cutoff $\varepsilon=k_{\text{min}}/k_{\text{max}}$ based on the CMB multipole range of interest, $l_{\text{min}}=2$ and $l_{\text{max}}=2000$, so that \cite{Fergusson:2008ra}
\begin{align}
    \varepsilon=\frac{k_{\text{min}}}{k_{\text{max}}}=\frac{l_{\text{min}}}{l_{\text{max}}}=10^{-3}.
\end{align}
We impose this cut-off on the integral when $r\to 0$.
\begin{figure}[t]
\centering
\includegraphics[scale=0.65]{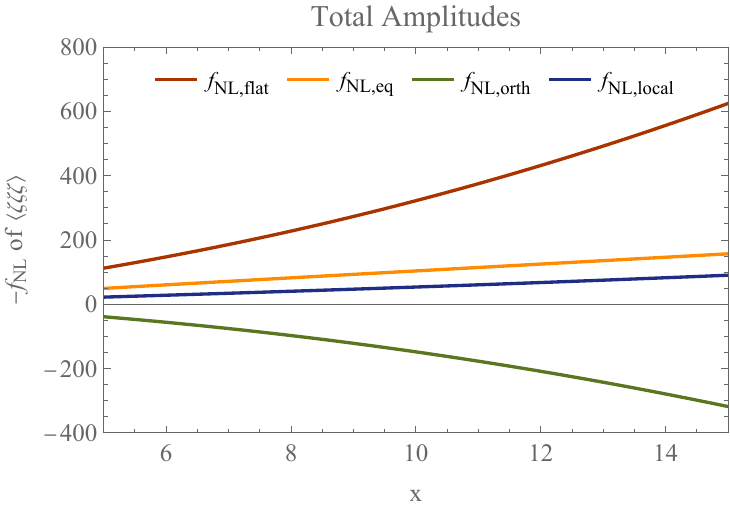}
\caption{\label{fig:Atot}Total projected amplitudes $-f_{\rm NL,S}$ of $\langle\zeta\zeta\zeta\rangle$ after restoring all coefficients $\Lambda_i$.}
\end{figure}

The correlations are shown in FIG.~\ref{fig:Ceq} and FIG.~\ref{fig:Clocal}. The $\dot{\mathcal{R}}^3$ and $\dot{\mathcal{R}}(\partial_i\mathcal{R})^2$ contributions, which also occur in multi-field inflation, are displayed in FIG.~\ref{fig:Ceq}. Both correlate strongly with the flattened template, at a level of order $0.9$, while their correlation with the local template is below $0.5$. Their flattened correlations decrease as $x$ grows because the overlap with other nonlocal templates increases. The $\dot{\mathcal{R}}^3$ shape remains predominantly flattened, evolving from a flattened-plus-equilateral mixture toward a flattened-plus-orthogonal one. The $\dot{\mathcal{R}}(\partial_i\mathcal{R})^2$ contribution gives an even cleaner flattened-plus-orthogonal mixture.

The $\mathcal{R}\dot{\mathcal{R}}^2$ interaction and the boundary term provide the two additional shapes in this model. Across the displayed range of $x$, $C(\mathcal{R}\dot{\mathcal{R}}^2,\text{local})\simeq C(\text{Boundary},\text{local})\simeq1$, so both contributions closely follow the local template. Their correlations with the remaining templates depend weakly on $x$ and remain below $0.5$. 

The amplitudes of the individual interactions are shown in FIG.~\ref{fig:Aflat} and FIG.~\ref{fig:Alocal}. The coefficients $\Lambda_i$ are not included in these figures because they are all of order $\mathcal{O}(10)$. For the $\dot{\mathcal{R}}^3$ and $\dot{\mathcal{R}}(\partial_i\mathcal{R})^2$ interactions, $f_{\text{NL}}^{\text{flat}}$ is the largest projected amplitude and grows rapidly as $x^3$. A nonzero local projection in this case reflects template overlap rather than a physically local squeezed limit; the growing equilateral projection has the same interpretation.

For the $\mathcal{R}\dot{\mathcal{R}}^2$ interaction, the $x$ dependence of the local amplitude is milder than that of the flattened projection: the former is linear in $x$, whereas the latter grows as $x^2$. Although $f_{\text{NL}}^{\text{flat}}$ is the largest projection, the shape itself is close to local. This is consistent because a projected amplitude depends on template normalization and non-orthogonal overlap, not solely on shape similarity.

The boundary term likewise has a non-vanishing local contribution. It remains subdominant, and all of its projected amplitudes are nearly independent of $x$, as expected.

We now combine the individual interactions into the total projected amplitude of $\langle\zeta\zeta\zeta\rangle$ for a template $S$:
\begin{align}
    f_{\text{NL},S}=-\left(\Lambda_1f_{\text{NL},\dot{\mathcal{R}}^3}^{\text{S}}+\Lambda_2f_{\text{NL},\dot{\mathcal{R}}(\partial_i\mathcal{R})^2}^{\text{S}}+\Lambda_3f_{\text{NL},\mathcal{R}\dot{\mathcal{R}}^2}^{\text{S}}+\Lambda_5f_{\text{NL},\text{bdy}}^{\text{S}}\right)+\Lambda_6f^{\rm S}_{\rm NL,nl}.
\end{align}
The minus sign comes from the minus sign of the first term in the (\ref{zeta3point}). These amplitudes are shown in FIG.~\ref{fig:Atot}, where all coefficients $\Lambda_i$ have been restored. As a result, for not too large $x$, the flattened shape is dominated by both the $x^2$- and $x^3$-enhanced contributions (the former comes from $\mathcal{R}\dot{\mathcal{R}}^2$ while the latter comes from $\dot{\mathcal{R}}(\partial_i\mathcal{R})^2$ and $\dot{\mathcal{R}}^3$), while for large $x$, the $x^3$-enhanced contribution takes over. The local-shape contribution, although enhanced by $x$, remains subdominant compared with the flattened and equilateral shapes.

\begin{figure}[t]
\centering
\includegraphics[scale=0.8]{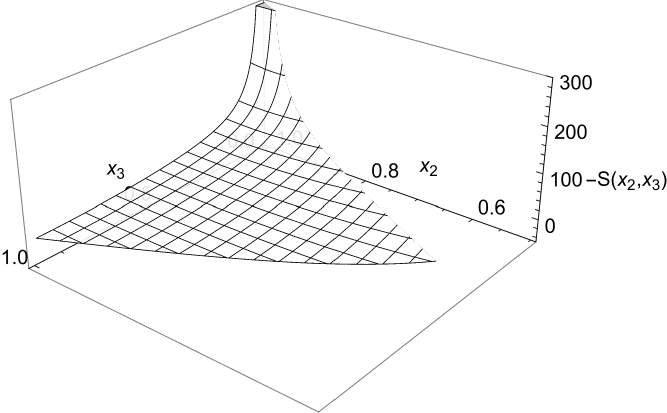}
\caption{\label{fig:Stot}Total shape function at $x=5$. The local shape is prominent.}
\end{figure}

\section{Conclusion}
We have studied primordial non-Gaussianities in an isotropic inflationary background supported by vector fields coupled to the inflaton. The parameter $h=\sqrt{\epsilon_A/2\epsilon_\phi}$ provides a useful separation between two qualitatively different regimes. For $h\ll1$, the vector perturbation is weakly coupled to the curvature mode, but its superhorizon transfer accumulates over time. The resulting local bispectrum scales as $h^2N_K^3$ and therefore strongly constrains the weak-vector contribution.

Our primary result concerns $h\gg1$. In this regime the entropic perturbation is heavy and can be integrated out, yielding a low-energy EFT for the curvature perturbation with an imaginary sound speed. The transient growth associated with this EFT leaves a characteristic imprint on the bispectrum. 
The different cubic operators generate a characteristic hierarchy of momentum-space signals. 
The interactions $\dot{\mathcal{R}}^{3}$ and $\dot{\mathcal{R}}(\partial_i\mathcal{R})^{2}$ 
produce flattened-enhanced bispectra whose amplitudes scale as $x^{3}$. 
The vector-field EFT additionally contains the interaction 
$\mathcal{R}\dot{\mathcal{R}}^{2}$, which generates a flattened contribution scaling as $x^{2}$ 
and a genuine squeezed contribution scaling as $x/r$, where $r$ is the ratio of the long to 
short wavelength. The temporal boundary term and the nonlinear transformation from 
$\mathcal{R}$ to the uniform-density curvature perturbation $\zeta$ provide further local 
contributions, but these are not enhanced by a positive power of $x$.

Our template analysis reflects this operator structure. The $\dot{\mathcal{R}}^{3}$ and 
$\dot{\mathcal{R}}(\partial_i\mathcal{R})^{2}$ shapes correlate strongly with the flattened 
template, whereas the $\mathcal{R}\dot{\mathcal{R}}^{2}$ and boundary contributions possess 
pronounced local components. After restoring the interaction coefficients and converting to 
the observable $\zeta$ bispectrum, the total signal is a nontrivial mixture of flattened, 
equilateral, orthogonal, and local projections. In the parameter range studied here, the 
flattened and equilateral projections dominate the total amplitude, while the local projection 
remains significant and grows linearly with $x$. The characteristic prediction of the 
vector-supported regime is therefore the coexistence of flattened signals scaling as $x^{3}$ 
and $x^{2}$ with a genuine local component scaling as $x$. The total shape function is shown in FIG.~\ref{fig:Stot}.

These results distinguish kinetically coupled vector-field inflation from scalar multifield 
realizations with an imaginary sound speed, for which the $x^{3}$-enhanced flattened operators 
are familiar but the vector-specific $\mathcal{R}\dot{\mathcal{R}}^{2}$ interaction supplies an 
additional squeezed signal. The simultaneous presence of these components provides a potentially 
useful observational discriminator even when the background is exactly isotropic.

In this work, we only consider an isotropic configuration of vector fields for computational simplicity. We emphasize again that this also applies to the case with a single vector field, and due to the strong anisotropy, the case with only one vector field is more interesting \cite{Chen:2025qyv}. We will discuss this in the future. One of the biggest differences between the case with vector field coupling and the case with scalar field coupling (multi-field inflation) is the non-Gaussianities with a non-negligible local shape. This non-Gaussianities indicate correlation between long wavelength and short wavelength modes. As studied in \cite{Chen:2023bcz}, the production of primordial black holes and induced gravitational waves on small scales \cite{Palma:2020ejf,Fumagalli:2020adf,Braglia:2020eai,Fumagalli:2020nvq,Fumagalli:2021cel}, and also an enhancement on primordial GWs \cite{Fumagalli:2021mpc} for EFT with imaginary sound speed are possible. The information of small-scale anisotropy of IGWs can be encoded into the large-scale anisotropy due to the non-trivial local shape non-Gaussianities generated by coupling with vector fields. It is interesting to explore the anisotropy of the stochastic gravitational wave background in the presence of kinetic coupling with vector fields for $h\gg1$ \cite{Chen:2022qec,Kuang:2023urj,Xie:2026xhr}. We also leave this discussion for future work.

\appendix
\section{Quadratic and cubic action}\label{appendix}
The perturbed Lagrangian up to third order is
\begin{align}
    \mathcal{L}\simeq{}&-\frac{1}{4}\sqrt{-g}g^{\mu\alpha}g^{\nu\beta}\left[f^2+(f^2)_{\phi}\delta\phi+\frac{1}{2!}(f^2)_{\phi\phi}\delta\phi^2+\frac{1}{3!}(f^2)_{\phi\phi\phi}\delta\phi^3\right]\nonumber\\
    &\times\left(F^{(a)}_{\mu\nu}+\delta F^{(a)}_{\mu\nu}\right)\left(F^{(a)}_{\alpha\beta}+\delta F^{(a)}_{\alpha\beta}\right),
\end{align}
The quadratic part is 
\begin{align}
    \mathcal{L}^{(2)}={}&\frac{a^3}{2}\delta\dot{\phi}^2-\frac{a^3}{2}\left(\frac{1}{a^2}k^2+V_{\phi\phi}\right)\delta\phi^2+\frac{3a}{2}f^2\delta\dot{A}^2-\frac{1}{a}f^2k^2\delta A^2+\frac{a}{2}f^2k^4Y^2+af^2k^2 Y\delta \dot{A}\nonumber\\
    &+2aff_{\phi}\dot{A}\left(3\delta\dot{A}+k^2Y\right)\delta\phi+\frac{3a}{2}\left(f_{\phi}^2+ff_{\phi\phi}\right)\dot{A}^2\delta\phi^2+af^2k^2\dot{U}^2-\frac{1}{a}f^2k^4U^2,
\end{align}
The perturbation $\mathbb{Y}$ is non-dynamical and can be eliminated through the quadratic action:
\begin{align}\label{eq5}
    \partial^2\mathbb{Y}=\delta\dot{\mathbb{A}}+\frac{(f^2)_{\phi}}{f^2}\dot{\mathbb{A}}\delta\phi,\ \ \ \ \ \ \partial_i\partial_j\mathbb{Y}=\frac{1}{3}\partial^2\mathbb{Y}\delta_{ij}.
\end{align}
The cubic part is
\begin{align}\label{L3}
    \mathcal{L}^{(3)}
\simeq{}&\frac{a^3}{4}\frac{(f^2)_{\phi\phi\phi}}{f^2}\frac{f^2\dot{\mathbb{A}}^2}{a^2}\delta\phi^3+\frac{a^3}{2}\frac{(f^2)_{\phi\phi}}{f^2}\frac{f^2\dot{\mathbb{A}}}{a^2}\left(3\delta\dot{\mathbb{A}}-\partial^2\mathbb{Y}\right)\delta\phi^2\nonumber\\
    &+\frac{a^3}{2}\frac{(f^2)_{\phi}}{f^2}\frac{f^2}{a^2}\left(3\delta\dot{\mathbb{A}}^2-\frac{2}{a^2}\partial_i\delta\mathbb{A}\partial_i\delta\mathbb{A}+\partial_i\partial_j\mathbb{Y}\partial_i\partial_j\mathbb{Y}-2\delta\dot{\mathbb{A}}\partial^2\mathbb{Y}\right)\delta\phi\nonumber\\
    &+a^3\frac{(f^2)_{\phi}}{f^2}\frac{f^2}{a^2}\left(\partial_i\dot{\mathbb{U}}\partial_i\dot{\mathbb{U}}-\frac{1}{2a^2}\partial_i\partial_j\mathbb{U}\partial_i\partial_j\mathbb{U}-\frac{1}{2a^2}\partial^2\mathbb{U}\partial^2\mathbb{U}\right)\delta\phi.
\end{align}
Using (\ref{eq2}), we obtain
\begin{align}\label{eq3}
    \frac{(f^2)_{\phi}}{f^2}\simeq\frac{2\sqrt{2}}{\sqrt{\epsilon_\phi}}\frac{1}{M_{\text{pl}}},\ \ \ \ \ \ \ \frac{(f^2)_{\phi\phi}}{f^2}\simeq\frac{8}{\epsilon_\phi}\frac{1}{M_{\text{pl}}^2},\ \ \ \ \ \ \ \frac{(f^2)_{\phi\phi\phi}}{f^2}\simeq\left(\frac{8}{\epsilon_\phi}\right)^{3/2}\frac{1}{M_{\text{pl}}^3},
\end{align}
where only the dominant terms in the slow-roll approximation have been retained. With these approximations, eliminating the non-dynamical perturbation $\mathbb{Y}$ yields the cubic action (\ref{L3phiQ}).

\subsection{Small $h$}
For small $h$, the cubic action can be expanded in powers of $h$. The leading-order terms are
\begin{align}\label{L3smallh}
    \mathcal{L}^{(3)}_{h\ll1}\simeq{}\frac{a^3}{M_{\text{pl}}\sqrt{2\epsilon}}\bigg[\mathcal{L}^{(3)}_0+h\mathcal{L}^{(3)}_1+h^2\mathcal{L}^{(3)}_2+\mathcal{O}(h^3)\bigg],
\end{align}
where
\begin{align}
    \mathcal{L}^{(3)}_0={}&
-\frac{4}{3}\mathcal R_c\dot{\mathcal F}^{2}
-8H\mathcal R_c\mathcal F\dot{\mathcal F}
-12H^2\mathcal R_c\mathcal F^2
+\frac{2}{a^2}\mathcal R_c\partial_i\mathcal F\partial_i\mathcal F-\Xi_U\mathcal{R}_c,\label{L3_0}\\
\mathcal{L}^{(3)}_1={}&\frac{8\sqrt2}{3}\mathcal R_c\dot{\mathcal R}_c\dot{\mathcal F}
+8\sqrt2 H\mathcal R_c\mathcal F\dot{\mathcal R}_c
+\frac{16\sqrt2}{3}H\mathcal R_c^2\dot{\mathcal F}
+16\sqrt2 H^2\mathcal R_c^2\mathcal F-\frac{4\sqrt2}{3}\mathcal F\dot{\mathcal F}^{2}
\nonumber\\
&-8\sqrt2 H\mathcal F^2\dot{\mathcal F}-12\sqrt2 H^2\mathcal F^3+\frac{2\sqrt2}{a^2}\mathcal F(\partial_i\mathcal F)^2
-\frac{4\sqrt2}{a^2}\mathcal R_c\partial_i\mathcal R_c\partial_i\mathcal F-\sqrt{2}\Xi_U\mathcal{F},\label{L3_1}\\
\mathcal{L}^{(3)}_2={}&-\frac{8}{3}H^2\mathcal R_c^3
-\frac{32}{3}H\mathcal R_c^2\dot{\mathcal R}_c
-\frac{8}{3}\mathcal R_c\dot{\mathcal R}_c^{2}+40H^2\mathcal R_c\mathcal F^2
+16H\mathcal F^2\dot{\mathcal R}_c
+\frac{64}{3}H\mathcal R_c\mathcal F\dot{\mathcal F}
\nonumber\\
&+\frac{16}{3}\mathcal F\dot{\mathcal R}_c\dot{\mathcal F}+\frac{8}{3}\mathcal R_c\dot{\mathcal F}^{2}+\frac{4}{a^2}\mathcal R_c(\partial_i\mathcal R_c)^2
-\frac{8}{a^2}\mathcal F\partial_i\mathcal R_c\partial_i\mathcal F
-\frac{4}{a^2}\mathcal R_c(\partial_i\mathcal F)^2.\label{L3_2}
\end{align}
Combining this result with the quadratic Lagrangian, we derive the conjugate momenta
\begin{align}
    a^{-3}\Pi_{\mathcal{R}_c}={}&\left( \dot{\mathcal{R}}_c - 4\sqrt{2} h H \mathcal{F} \right)
+ \frac{1}{\sqrt{2\epsilon}M_{\text{pl}}} \Bigg[ h \left( \frac{8\sqrt{2}}{3} \mathcal{R}_c \dot{\mathcal{F}} + 8\sqrt{2} H \mathcal{R}_c \mathcal{F} \right)\nonumber\\
&+ h^2 \left( -\frac{32}{3} H \mathcal{R}_c^2 - \frac{16}{3} \mathcal{R}_c \dot{\mathcal{R}}_c + 16H \mathcal{F}^2 + \frac{16}{3} \mathcal{F} \dot{\mathcal{F}} \right) \Bigg],\\
a^{-3}\Pi_{\mathcal{F}} ={}& \dot{\mathcal{F}}+ \frac{1}{\sqrt{2\epsilon}M_{\text{pl}}} \Bigg[-\frac{8}{3}\mathcal{R}_c \dot{\mathcal{F}} - 8H\mathcal{R}_c\mathcal{F}+ h \bigg( \frac{8\sqrt{2}}{3}\mathcal{R}_c \dot{\mathcal{R}}_c + \frac{16\sqrt{2}}{3}H\mathcal{R}_c^2 \nonumber\\
& - \frac{8\sqrt{2}}{3}\mathcal{F} \dot{\mathcal{F}} - 8\sqrt{2}H\mathcal{F}^2 \bigg)+ h^2 \left( \frac{64}{3}H\mathcal{R}_c\mathcal{F} + \frac{16}{3}\mathcal{F} \dot{\mathcal{R}}_c + \frac{16}{3}\mathcal{R}_c \dot{\mathcal{F}} \right) \Bigg],\\
a^{-3}\Pi_{\mathbb{U}}^i={}&\frac{2f^2}{a^2}\partial_i\dot{\mathbb{U}}-\frac{8\sqrt{2}}{\sqrt{2\epsilon}M_{\text{pl}}}\frac{f^2}{a^2}\left(\mathcal{R}_c+\sqrt{2}h\mathcal{F}\right)\partial_i\dot{\mathbb{U}}.
\end{align}
We take the $h=0$ diagonal quadratic Lagrangian as the free theory, and treat the rest of the other terms as interactions. The Hamiltonian density is
$\mathcal{H}_I=\Pi_{\mathcal{R}_c}\dot{\mathcal{R}}_c+\Pi_{\mathcal{F}}\dot{\mathcal{F}}+\Pi_{\mathbb{U}}^i\partial_i\dot{\mathbb{U}}-\mathcal{L}$. Inverting the momentum–velocity relation perturbatively by using conjugate momentum, we obtain the interacting Hamiltonian as
\begin{align}\label{H_smallh}
    H_{I,h\ll 1}=\int d^3x{}\left(\mathcal{H}_I^{(2)}+\mathcal{H}_I^{(3)}\right).
\end{align}
The interaction Hamiltonian contains two parts. Because of the derivative mixing $\mathcal{F}\dot{\mathcal{R}}_c$, its quadratic and cubic terms are not simply $-\mathcal{L}^{(2)}_{\text{mix}}$ and $-\mathcal{L}^{(3)}_{h\ll1}$. Instead, we find
\begin{align}
    \mathcal{H}_I^{(2)}={}&-\mathcal{L}^{(2)}_{\text{mix}}+\frac{1}{2a^3}\left(\frac{\partial\mathcal{L}^{(2)}_{\text{mix}}}{\partial\dot{\mathcal{R}}_c}\right)^2=a^3\left(4\sqrt{2}hH\mathcal{F}\dot{\mathcal{R}}_c+8h^2H^2\mathcal{R}_c^2+4h^2H^2\mathcal{F}^2\right),\\
    \mathcal{H}_I^{(3)}={}&-\mathcal{L}^{(3)}_{h\ll1}+\frac{1}{a^3}\frac{\partial\mathcal{L}^{(2)}_{\text{mix}}}{\partial\dot{\mathcal{R}}_c}\frac{\partial\mathcal{L}^{(3)}_{h\ll1}}{\partial\dot{\mathcal{R}}_c}=-\mathcal{L}^{(3)}_{h\ll1}-4\sqrt{2}hH\mathcal{F}\frac{\partial\mathcal{L}^{(3)}_{h\ll1}}{\partial\dot{\mathcal{R}}_c},
\end{align}
where $\mathcal{L}^{(2)}_{\text{mix}}=-4\sqrt{2}a^3hH\mathcal{F}\dot{\mathcal{R}}_c+\mathcal{O}(h^3)$ and $\mathcal{L}^{(3)}_{h\ll1}$ is given by (\ref{L3smallh}).

\subsection{Large $h$}
For large $h$, the heavy field can be integrated out by inserting the leading solution of $\mathcal{F}$ into the cubic Lagrangian. We can expand the solution as
\begin{align}
    \mathcal{F}=\frac{1}{m_s^2}\sum^{\infty}_{i=0}\left(\frac{\square}{m_s^2}\right)^{i}\big(-\sqrt{2}hH\big)\left(4\dot{\mathcal{R}}_c+\frac{16h^2}{1+2h^2}H\mathcal{R}_c\right).
\end{align}
The leading order solution is given by (\ref{FLO}). In this appendix we keep the next-to-leading order contribution of this solution, i.e., 
\begin{align}
    \mathcal{F}=\mathcal{F}_{\rm LO}\left[1+\delta_{\square}+\mathcal{O}\left(\frac{\square^2}{m_s^4}\right)\right],\ \ \ \ \ \delta_{\square}\equiv\frac{1}{m_s^2}\frac{\square\mathcal{F}_{\rm LO}}{\mathcal{F}_{\rm LO}}.
\end{align}
We can expand the cubic Lagrangian in the $\delta_{\square}\ll1$ limit. This gives the effective quadratic Lagrangian
\begin{align}
    \mathcal{L}^{(2)}_{\text{eff}}=\frac{a^3}{2}\Bigg[\bigg(1-\frac{4\left(1+2h^2\right)}{3+2h^2}\bigg)\dot{\mathcal{R}}_c^2-\frac{1}{a^2}(\partial_i \mathcal{R}_c)^2\Bigg]+\Delta\mathcal{L}^{(2)}_{\rm eff},
\end{align}
where the correction of quadratic Lagrangian is
\begin{align}
    \Delta\mathcal{L}^{(2)}_{\rm eff}=-a^3
\left[
\frac{2(1+2h^2)}{3+2h^2}
\dot{\mathcal{R}}_c^{\,2}
+
\frac{16h^2H}{3+2h^2}
\mathcal{R}_c\dot{\mathcal{R}}_c
+
\frac{32h^4H^2}
{(1+2h^2)(3+2h^2)}
\mathcal{R}_c^{\,2}
\right]\delta_{\square}.
\end{align}
We have shown in the FIG. \ref{fig:delta}(left) that when the EFT is valid, the mode is still inside the horizon. In this case, we can take $\delta_{\square}\simeq \partial^2/m_s^2$. Under this approximation, the Lagrangian can be rewritten via integration by parts as
\begin{align}
    \mathcal{L}^{(2)}_{\text{eff}}=\frac{a^3}{2c_{s,\mathrm{eff}}^2}
\left[
\dot{\mathcal{R}}_c^{\,2}
-
c_{s,\mathrm{eff}}^2\,
(\partial_i\mathcal{R}_c)^{2}
\right],
\end{align}
where
\begin{align}
    c_{s,\mathrm{eff}}^2
{}&=
c_s^2
+
\frac{4(1+2h^2)(3+2h^2)}{(1+6h^2)^2}
\delta_{\square}+\mathcal{O}\left(\delta_{\square}^2\right)\nonumber\\
&\simeq{c_s^2}\left(1-\frac{4}{3}\delta_{\square}\right)+\mathcal{O}\left(\frac{\delta_{\square}}{h^2},\frac{1}{h^4},\delta_{\square}^2\right)
\end{align}
and the leading-order of the speed of sound $c_s$ is defined by (\ref{cs}). In the quadratic Lagrangian, with $A=0.3$, we finds $\delta_{\square}\sim0.1$, implying that the speed of sound receives a correction of only about $10\%$.

Similarly, the cubic Lagrangian is given by
\begin{align}\label{L3firstorder}
    \mathcal{L}^{(3)}_{\text{eff}}={}&\frac{a^3}{M_{\text{pl}}\sqrt{2\epsilon}}\bigg[108H^2\mathcal{R}_c^3+108H\mathcal{R}_c^2\dot{\mathcal{R}}_c+36\mathcal{R}_c\dot{\mathcal{R}}_c^2+\frac{4}{H}\dot{\mathcal{R}}_c^3+\frac{6}{a^2}\mathcal{R}_c(\partial_i\mathcal{R}_c)^2\nonumber\\
    &+\frac{2}{a^2H}\dot{\mathcal{R}}_c(\partial_i\mathcal{R}_c)^2\bigg]+\mathcal{L}_{\delta_{\square}}^{(3)}+\mathcal{L}_{1/h^{2}}^{(3)}+\mathcal{O}\left(\frac{\delta_{\square}}{h^2},\frac{1}{h^4},\delta_{\square}^{2}\right),
\end{align}
where
\begin{align}
    \mathcal{L}_{\delta_{\square}}^{(3)}={}&\frac{a^3}{M_{\mathrm{pl}}\sqrt{2\epsilon}}
\Bigg[
216H^2\mathcal R_c^3
+252H\mathcal R_c^2\dot{\mathcal R}_c
+96\mathcal R_c\dot{\mathcal R}_c^{\,2}
+\frac{12}{H}\dot{\mathcal R}_c^{\,3}+\frac{4}{a^2}\mathcal R_c(\partial_i\mathcal R_c)^2\nonumber\\
&+\frac{2}{a^2H}\dot{\mathcal R}_c
(\partial_i\mathcal R_c)^2
\Bigg]\delta_\square,\\
\mathcal{L}_{1/h^2}^{(3)}={}&\frac{a^3}{M_{\mathrm{pl}}\sqrt{2\epsilon}}\Bigg[
378H^2\mathcal R_c^3
+378H\mathcal R_c^2\dot{\mathcal R}_c
+126\mathcal R_c\dot{\mathcal R}_c^{\,2}
+\frac{14}{H}\dot{\mathcal R}_c^{\,3}+\frac{21}{a^2}\mathcal R_c(\partial_i\mathcal R_c)^2
\\
&+\frac{7}{a^2H}\dot{\mathcal R}_c
(\partial_i\mathcal R_c)^2
+\left(
\frac{6}{a^2H}\mathcal R_c
+\frac{2}{a^2H^2}\dot{\mathcal R}_c
\right)
\partial_i\mathcal R_c\,
\partial_i\dot{\mathcal R}_c
\Bigg]\frac{1}{h^2}.
\end{align}
We have also discarded all the terms $H\dot{\mathcal{F}}_{\rm LO}\delta_{\square}\lesssim\mathcal{O}(\square^2/m_s^4)$ and magnetic perturbations $\mathbb{U}$ in the Lagrangian. We find that the term $\dot{\delta}_{\square}\sim1/h^4$, therefore, does not appear in the formula. It is clear to see the corrections associated with $\square/m_s^2$ and $1/h^2$ separately. The corrections from $\delta_{\square}$ are
\begin{align}
    &\frac{\Delta\mathcal{L}^{(3)}_{\mathcal{R}^3}}{\mathcal{L}^{(3)}_{\mathcal{R}^3}}\sim 2\delta_{\square},\ \ \ \ \ \ \ \ \ \frac{\Delta\mathcal{L}^{(3)}_{\mathcal{R}^2\dot{\mathcal{R}}}}{\mathcal{L}^{(3)}_{\mathcal{R}^2\dot{\mathcal{R}}}}\sim 2.3\delta_{\square},\ \ \ \ \ \ \ \ \ \frac{\Delta\mathcal{L}^{(3)}_{\mathcal{R}\dot{\mathcal{R}}^2}}{\mathcal{L}^{(3)}_{\mathcal{R}\dot{\mathcal{R}}^2}}\sim2.6\delta_{\square},\nonumber\\
    &\frac{\Delta\mathcal{L}^{(3)}_{\dot{\mathcal{R}}^3}}{\mathcal{L}^{(3)}_{\dot{\mathcal{R}}^3}}\sim3\delta_{\square},\ \ \ \ \ \ \ \ \ \frac{\Delta\mathcal{L}^{(3)}_{\mathcal{R}(\partial_i\mathcal{R})^2}}{\mathcal{L}^{(3)}_{\mathcal{R}(\partial_i\mathcal{R})^2}}\sim0.7\delta_{\square},\ \ \ \ \ \frac{\Delta\mathcal{L}^{(3)}_{\dot{\mathcal R}
(\partial_i\mathcal R)^2}}{\mathcal{L}^{(3)}_{\dot{\mathcal R}
(\partial_i\mathcal R)^2}}\sim \delta_{\square}.
\end{align}
On the other hand, $1/h^2\lesssim0.04$ for $h\gtrsim5$, which is small enough so that we can neglect these contributions. If we discard all the terms $1/h^2$ and $\delta_{\square}$, we can finally obtain the leading cubic action
\begin{align}
    \mathcal{L}^{(3)}_{\text{eff}}\simeq{}&\frac{a^3}{M_{\text{pl}}\sqrt{2\epsilon}}\bigg[108H^2\mathcal{R}_c^3+108H\mathcal{R}_c^2\dot{\mathcal{R}}_c+36\mathcal{R}_c\dot{\mathcal{R}}_c^2+\frac{4}{H}\dot{\mathcal{R}}_c^3+\frac{6}{a^2}\mathcal{R}_c(\partial_i\mathcal{R}_c)^2\nonumber\\
    &+\frac{2}{a^2H}\dot{\mathcal{R}}_c(\partial_i\mathcal{R}_c)^2-2\sqrt{2}\frac{f^2}{a^2}\left(\partial_i \dot{\mathbb{U}}\partial_i\dot{\mathbb{U}}-\frac{1}{2a^2}\partial_i\partial_j\mathbb{U}\partial_i\partial_j\mathbb{U}-\frac{1}{2a^2}\partial^2\mathbb{U}\partial^2 \mathbb{U}\right)\nonumber\\
    &\times\left(3\mathcal{R}_c+\frac{1}{H}\dot{\mathcal{R}}_c\right)\bigg],
\end{align}
where we have also recovered the magnetic perturbation $\mathbb{U}$. We can continue to simplify this cubic action. In cubic action, the second term $\mathcal{R}_c^2\dot{\mathcal{R}}_c$ can be rewritten as
\begin{align}
    \frac{108a^3H}{M_{\text{pl}}\sqrt{2\epsilon}}\mathcal{R}_c^2\dot{\mathcal{R}}_c=\partial_t\left(\frac{108a^3H}{M_{\text{pl}}\sqrt{2\epsilon}}\mathcal{R}_c^3\right)-\frac{108a^3H^2}{M_{\text{pl}}\sqrt{2\epsilon}}\mathcal{R}_c^3.
\end{align}
Using the equation of motion of $\mathcal{R}$ of the EFT, the third term $\mathcal{R}_c\dot{\mathcal{R}}_c^2$ can be rewritten as
\begin{align}
    \frac{36a^3}{M_{\text{pl}}\sqrt{2\epsilon}}\mathcal{R}_c\dot{\mathcal{R}}_c^2
    ={}&\partial_t\left(\frac{18a^3\mathcal{R}_c^2\dot{\mathcal{R}}_c}{M_{\text{pl}}\sqrt{2\epsilon}}\right)-\partial_i\left(\frac{18ac_s^2}{M_{\text{pl}}\sqrt{2\epsilon}}\mathcal{R}_c^2\partial_i\mathcal{R}_c\right)+\frac{36ac_s^2}{M_{\text{pl}}\sqrt{2\epsilon}}\mathcal{R}_c\partial_i\mathcal{R}_c\partial_i\mathcal{R}_c.
\end{align}
Thus the term $\mathcal{R}_c\partial_i\mathcal{R}_c\partial_i\mathcal{R}_c$ in the cubic action can be replaced by $\mathcal{R}_c\dot{\mathcal{R}}_c^2$ and a temporal boundary term, with spatial boundary terms neglected. This procedure gives (\ref{L3fin}), and the corresponding interaction Hamiltonian is given by $H^{(3)}_{I,\ \text{eff}}=-\int d^3x\mathcal{L}^{(3)}_{\text{eff}}$.

\section{Estimation of corrections to bispectrum}\label{appendix2}
To estimate the contribution from the correction operators, we need to compute both the $\delta_{\square}$ correction and the $1/h^2$ correction. Since we already know that the $1/h^2$ correction is sufficiently small in the parameter region of interest, we focus here only on the $\delta_{\square}$ correction. This correction becomes small enough by the time the mode crosses the horizon, so we only evaluate its contribution while the mode is still inside the horizon, where we may take
\begin{align}
    \delta_{\square}\sim-\frac{p_m^2}{m_s^2}=-\frac{H^2k_m^2}{m_s^2}\eta^2,
\end{align}
where $p_m\equiv k_m/a$. After integration by parts, we obtain
\begin{align}
    \mathcal{L}_{\delta_{\square}}^{(3)}
={}&
\frac{a^3}{M_{\mathrm{pl}}\sqrt{2\epsilon}}
\left[
\frac{12}{H}\dot{\mathcal{R}}_c^{\,3}
+
\frac{2}{a^2H}
\dot{\mathcal{R}}_c
\left(
\partial_i\mathcal{R}_c
\right)^2
+
84\mathcal{R}_c\dot{\mathcal{R}}_c^{\,2}
+
128H^2\mathcal{R}_c^{\,3}
\right]\delta_{\square}\nonumber
\\
&+
\frac{1}{M_{\mathrm{pl}}\sqrt{2\epsilon}}
\frac{d}{dt}
\left\{
a^3
\left[
88H\mathcal{R}_c^{\,3}
+
6\mathcal{R}_c^{\,2}\dot{\mathcal{R}}_c
\right]\delta_{\square}\right\}.
\end{align}
We compute the contributions arising solely from the operator corrections, under the assumption that the quadratic action is not modified. We here discuss the flattened non-Gaussianities as an example.

\textbf{Interaction in $\dot{\mathcal{R}}^3$: } The correction to the correlation from $\dot{\mathcal{R}}^3$ is given by
\begin{align}
    \Delta \langle\mathcal{R}_{\boldsymbol{k}_1}\mathcal{R}_{\boldsymbol{k}_2}\mathcal{R}_{\boldsymbol{k}_3}\rangle
    \supset{}&-\frac{H^2k_m^2}{m_s^2}\frac{864}{|c_s|^2\tilde{k}_1^2}|c_s|^2(2\pi^2)^2\mathcal{P}_{\mathcal{R}}^2(2\pi)^3\delta^{(3)}(\boldsymbol{k}_1+\boldsymbol{k}_2+\boldsymbol{k}_3)\frac{1}{(k_1k_2k_3)^2}\nonumber\\
    &\times\Bigg\{-\frac{k_1k_2k_3}{(k_1+k_2+k_3)^3}
    +\frac{k_1k_2k_3}{\tilde{k}_1^3}\bigg[1-e^{-\tilde{k}_1x/k_m}\bigg(1+\frac{\tilde{k}_1x}{k_m}+\frac{\tilde{k}_1^2x^2}{2k_m^2}\nonumber\\
    &+\frac{\tilde{k}_1^3x^3}{6k_m^3}+\frac{\tilde{k}_1^4x^4}{24k_m^4}\bigg)\bigg]\Bigg\}+\text{2 perm.}
\end{align}
In the flattened limit, we obtain
\begin{align}
    \Delta S_{\dot{\mathcal{R}}^3}(k_2+k_3,k_2,k_3)\simeq\frac{9}{5}\frac{H^2}{|m_s|^2}\frac{k_2k_3}{k_1^2}x^5.
\end{align}
The error is then given by
\begin{align}
    \frac{\Delta S_{\dot{\mathcal{R}}^3}}{S_{\dot{\mathcal{R}}^3}}(k_2+k_3,k_2,k_3)\simeq\frac{9}{5}\frac{H^2x^2}{|c_s|^2|m_s|^2}\sim\frac{9}{5}\delta_{\square},
\end{align}
where we have used $p_i^2/|m_s|^2\sim\delta_{\square}$.

\textbf{Interaction in $\dot{\mathcal{R}}(\partial_i\mathcal{R})^2$: } The correction to the correlation from $\dot{\mathcal{R}}(\partial_i\mathcal{R})^2$ is given by
\begin{align}
    \Delta \langle\mathcal{R}_{\boldsymbol{k}_1}\mathcal{R}_{\boldsymbol{k}_2}\mathcal{R}_{\boldsymbol{k}_3}\rangle
    \supset{}&\frac{H^2k_m^2}{m_s^2|c_s|^2}\,
(2\pi^2)^2\mathcal{P}_{\mathcal R}^{\,2}
(2\pi)^3
\delta^{(3)}
\left(
\mathbf{k}_1+\mathbf{k}_2+\mathbf{k}_3
\right)
\frac{1}{(k_1k_2k_3)^2}\nonumber\\
&\times
\Bigg\{
-\frac{k_1k_2k_3}{k_t^5}
\,12p_0(k_1,k_2,k_3)
+
\frac{k_1k_2k_3}{\widetilde{k}_1^{\,5}}
\Bigg[
12p_0(-k_1,k_2,k_3)\nonumber\\
&-
e^{-\widetilde{k}_1x/k_m}
\Bigg(
12p_0(-k_1,k_2,k_3)+
6\Big[
p_0(-k_1,k_2,k_3)
+p_1(-k_1,k_2,k_3)
\Big]
\frac{\widetilde{k}_1x}{k_m}
\nonumber\\
&+
2\Big[
p_0(-k_1,k_2,k_3)
+4p_1(-k_1,k_2,k_3)
+p_2(-k_1,k_2,k_3)
\Big]
\frac{\widetilde{k}_1^{\,2}x^2}{2k_m^2}\nonumber
\\
&+
6\Big[
p_1(-k_1,k_2,k_3)
+p_2(-k_1,k_2,k_3)
\Big]
\frac{\widetilde{k}_1^{\,3}x^3}{6k_m^3}+
12p_2(-k_1,k_2,k_3)
\frac{\widetilde{k}_1^{\,4}x^4}{24k_m^4}
\Bigg)
\Bigg]
\Bigg\}\nonumber
\\
&
+2~\mathrm{perm.}
\end{align}
In the flatten limit, we obtain
\begin{align}
    \Delta S_{\dot{\mathcal{R}}(\partial_i\mathcal{R})^2}(k_2+k_3,k_2,k_3)\simeq\frac{3}{10}\frac{H^2}{|c_s|^2|m_s|^2}\frac{k_2k_3}{k_1}x^5
\end{align}
The error is then given by
\begin{align}
    \frac{\Delta S_{\dot{\mathcal{R}}(\partial_i\mathcal{R})^2}}{S_{\dot{\mathcal{R}}(\partial_i\mathcal{R})^2}}(k_2+k_3,k_2,k_3)\simeq\frac{3}{5}\frac{H^2x^2}{|c_s|^2|m_s|^2}\sim\frac{3}{5}\delta_{\square}.
\end{align}

\textbf{Interaction in $\mathcal{R}\dot{\mathcal{R}}^2$: } The correction to the correlation from $\mathcal{R}\dot{\mathcal{R}}^2$ is given by
\begin{align}
    \Delta \langle\mathcal{R}_{\boldsymbol{k}_1}\mathcal{R}_{\boldsymbol{k}_2}\mathcal{R}_{\boldsymbol{k}_3}\rangle
    \supset{}&-84\,
\frac{H^2k_m^2}{m_s^2}\,
(2\pi^2)^2\mathcal{P}_{\mathcal R}^{\,2}
(2\pi)^3
\delta^{(3)}
\left(
\mathbf{k}_1+\mathbf{k}_2+\mathbf{k}_3
\right)
\frac{1}{(k_1k_2k_3)^2}\nonumber
\\
&\Bigg\{
\frac{k_1k_2k_3}{k_t^5}
\,12q_0(k_1,k_2,k_3)
-
\frac{k_1k_2k_3}{\widetilde{k}_1^{\,5}}
\Bigg[
12q_0(-k_1,k_2,k_3)\nonumber
\\
&-
e^{-\widetilde{k}_1x/k_m}
\Bigg(
12q_0(-k_1,k_2,k_3)+
6\Big[
q_0(-k_1,k_2,k_3)
+q_1(-k_1,k_2,k_3)
\Big]
\frac{\widetilde{k}_1x}{k_m}\nonumber
\\
&+
2\Big[
q_0(-k_1,k_2,k_3)
+4q_1(-k_1,k_2,k_3)
\Big]
\frac{\widetilde{k}_1^{\,2}x^2}{2k_m^2}+
6q_1(-k_1,k_2,k_3)
\frac{\widetilde{k}_1^{\,3}x^3}{6k_m^3}
\Bigg)
\Bigg]
\Bigg\}\nonumber
\\
&
+2~\mathrm{perm.}
\end{align}
In the flatten limit, we obtain
\begin{align}
    \Delta S_{\mathcal{R}\dot{\mathcal{R}}^2}(k_2+k_3,k_2,k_3)\simeq-\frac{21}{4}\frac{H^2}{|m_s|^2}\frac{k_2^2+k_2k_3+k_3^2}{k_1^2}x^4
\end{align}
We note that the correction terms have the same sign as the dominant term. The error is then given by
\begin{align}
    \frac{\Delta S_{\mathcal{R}\dot{\mathcal{R}}^2}}{S_{\mathcal{R}\dot{\mathcal{R}}^2}}(k_2+k_3,k_2,k_3)\simeq\frac{7}{3}\frac{H^2x^2}{|c_s|^2|m_s|^2}\sim\frac{7}{3}\delta_{\square}.
\end{align}
\begin{figure}[t]
\centering
\includegraphics[scale=0.6]{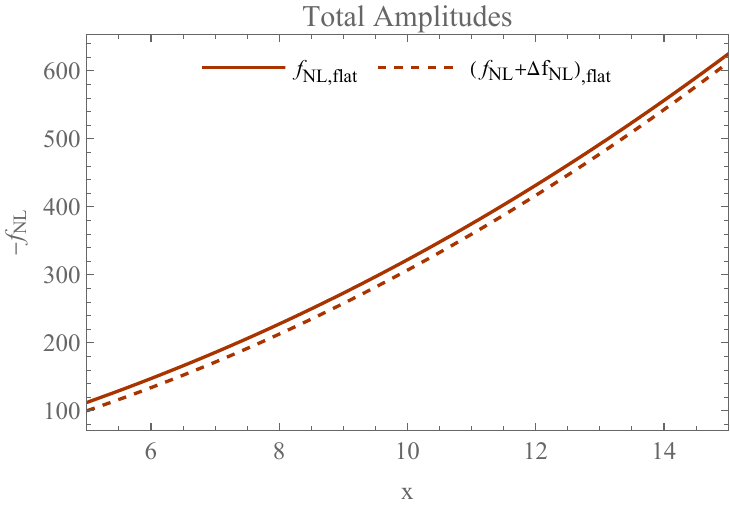}
\caption{\label{fig:fNLcorr}Effect of higher-derivative corrections on the total flattened 
non-Gaussianity. The solid curve shows the leading-order prediction 
for $f_{\rm NL}^{\rm flat}$, while the dashed curve includes the 
higher-derivative corrections.}
\end{figure}

\textbf{Interaction in $\mathcal{H}^{(3)}_{\text{bdy}}$: } The correction to the correlation from boundary is given by
\begin{align}
    \Delta \langle\mathcal{R}_{\boldsymbol{k}_1}\mathcal{R}_{\boldsymbol{k}_2}\mathcal{R}_{\boldsymbol{k}_3}\rangle
    \supset{}&6\,
\frac{H^2x^2}{m_s^2}\,
(2\pi^2)^2\mathcal{P}_{\mathcal R}^{\,2}
(2\pi)^3
\delta^{(3)}
\left(
\mathbf{k}_1+\mathbf{k}_2+\mathbf{k}_3
\right)
\frac{1}{(k_1k_2k_3)^2}\nonumber
\\
&\times
e^{-\widetilde{k}_1x/k_m}
\Bigg[
-\frac{k_m}{\widetilde{k}_1x}
l_{-1}(-k_1,k_2,k_3)
-l_0(-k_1,k_2,k_3)
-\frac{\widetilde{k}_1x}{k_m}
\Bigg]\nonumber
\\
&
+2~\mathrm{perm.}
\end{align}
In the flatten limit, we obtain
\begin{align}
    \Delta S_{\rm bdy}(k_2+k_3,k_2,k_3)\simeq\frac{3}{2}\frac{H^2}{|m_s|^2}\left[-2\left(\frac{k_1^2}{k_2k_3}+2 \text{ perm.}\right)+3+\frac{2}{x}\frac{k_2^2+k_2k_3+k_3^2}{k_2k_3}\right]x^2
\end{align}
The error is then given by
\begin{align}
    \frac{\Delta S_{\rm bdy}}{S_{\rm bdy}}(k_2+k_3,k_2,k_3)\simeq\frac{2}{3}\frac{H^2x^2}{|c_s|^2|m_s|^2}\sim\frac{2}{3}\delta_{\square}.
\end{align}
\textbf{Non-linear contribution: }The correction to the non-linear terms is given by
\begin{align}
    \Delta\zeta^{(2)}=\frac{12h^2(8h^2+9)}{(3+2h^2)^2}\delta_{\square}\mathcal{R}^2
\end{align}

We note that there is also an operator $\mathcal{R}^3$ that does not appear in the leading term. Nevertheless, owing to $\dot{\mathcal{R}}\sim p\mathcal{R}$ so that $\mathcal{R}^3/(\mathcal{R}\dot{\mathcal{R}}^2)\sim H^2/p^2\ll1$ inside the horizon, this contribution is subleading in the corrections, and thus we neglect it. In summary, the contributions from the higher-derivative corrections redshift the coefficients as
\begin{align}
    &\Lambda_{1}\to\Lambda_1\left(1+\frac{9}{5}\delta_{\square} \right),\ \ \ \ \ \ \Lambda_{2}\to\Lambda_2\left(1+\frac{3}{5}\delta_{\square} \right),\ \ \ \ \ \Lambda_{3}\to\Lambda_3\left(1+\frac{7}{3}\delta_{\square} \right),\nonumber\\ &\Lambda_{5}\to\Lambda_5\left(1+\frac{2}{3}\delta_{\square} \right),\ \ \ \ \ \Lambda_{6}\to\Lambda_6\left(1+\frac{48}{13}\delta_{\square} \right)
\end{align}
FIG. \ref{fig:fNLcorr} presents a comparison of the total $f_{\rm NL}^{\rm flat}$ with and without the corrections for $\delta_{\square}=0.05$. The relative error is found to be of the same order of magnitude as $\delta_{\square}$.

\acknowledgments
We would like to thank Meng-He Wu for his help with this work. We would also like to thank Zizhao He for providing us with delicious coffee beans during the completion of this work. This work was partly supported by  National Natural Science Foundation of China (NSFC) under Grants No. 12375049, Key Program of the Natural Science Foundation of Jiangxi Province under Grant No. 20232ACB201008, and the Ganpo High-Level Innovative Talent Program.


\bibliographystyle{JHEP}
\bibliography{biblio.bib}






\end{document}